\documentclass[pdflatex,sn-mathphys]{sn-jnl}

\usepackage{graphicx}
\usepackage{subcaption}

\algblockdefx[kernel]{BKernel}{EKernel}{\textbf{kernel }}{\textbf{end kernel}}

\usepackage{listings}

\jyear{2023}%

\usepackage{booktabs}

\usepackage{xcolor}
\usepackage{soul}

\definecolor{fondob}{RGB}{255, 0, 255} 
\definecolor{letrag}{RGB}{255, 255, 255} 

\theoremstyle{thmstyleone}%
\theoremstyle{thmstyletwo}%

\theoremstyle{thmstylethree}%
\newtheorem{definition}{Definition}%

\usepackage{subcaption}

\algdef{SE}[DOWHILE]{Do}{doWhile}{\algorithmicdo}[1]{\algorithmicwhile\ #1}%

\begin{document}

\title[GPolylla]{GPolylla: Fully GPU-accelerated polygonal mesh generator}


\author*[1]{\fnm{Sergio} \sur{Salinas-Fernández}}\email{ssalinas@dcc.uchile.cl}
\author[1]{\fnm{Roberto} \sur{Carrasco} \dgr{}}\email{rocarra@dcc.uchile.cl}
\author[1]{\fnm{Nancy} \sur{Hitschfeld-Kahler} \dgr{} }\email{nancy@dcc.uchile.cl}
\affil*[1]{\orgdiv{Department of Computer Sciences}, \orgname{Universidad de Chile}, \orgaddress{\\\street{Av. Beauchef 851}, \city{Santiago}, \postcode{8370456}, \state{RM}, \country{Chile}}}


\abstract{
This work presents a fully GPU-accelerated algorithm for the polygonal mesh generator known as Polylla. Polylla is a tri-to-polygon mesh generator, which benefits from the half-edge data structure to manage any polygonal shape. The proposed parallel algorithm introduces a novel approach to modify triangulations to get polygonal meshes using the half-edge data structure in parallel on the GPU. By changing the adjacency values of each half-edge, the algorithm accomplish to unlink half-edges that are not used in the new polygonal mesh without the need neither removing nor allocating new memory in the GPU. The experimental results show a speedup, reaching up to $\times 83.2$  when compared to the CPU sequential implementation. Additionally, the speedup is $\times 746.8$  when the cost of copying the data structure from the host device and back is not included.

}

\maketitle

\section{Introduction}
\label{sec:intro}

Polygonal mesh generation is becoming increasingly important due to the development of new numerical methods, such as the Virtual Element Method (VEM)~\cite{Vemapplications, Sorgente2022}. Meshes based on triangles and quadrangles are common in simulations using FEM\cite{FEM} to solve problems related to heat transfer \cite{FEMheattransfer}, fluid dynamics \cite{FemFluiddunamics} and fracture mechanics \cite{FEMfracturemechanics}, among others. However, FEM elements must obey specific quality criteria~\cite{KNUPP2003217}, such as having no excessively large obtuse angles or tiny angles, sides of graded length (aspect ratio criterion), and so on. To meet these requirements, it is sometimes necessary to insert a large number of points and elements to model a domain, which can increase the simulation time.

VEM, on the other hand, can use any polygon as a basic cell, improving the simulation speed because of the use of fewer cells than if only triangles and quadrilaterals are used to solve the same problem. Most VEM implementations currently use polygonal meshes formed by Voronoi cells~\cite{Voronoi}, which are convex polygons. Voronoi-based meshes work properly with the VEM but these meshes prevent the researchers or engineers from exploring the full potential of VEM on arbitrary polygonal meshes.

\begin{figure}
  \centering
\includegraphics[width=0.7\linewidth]{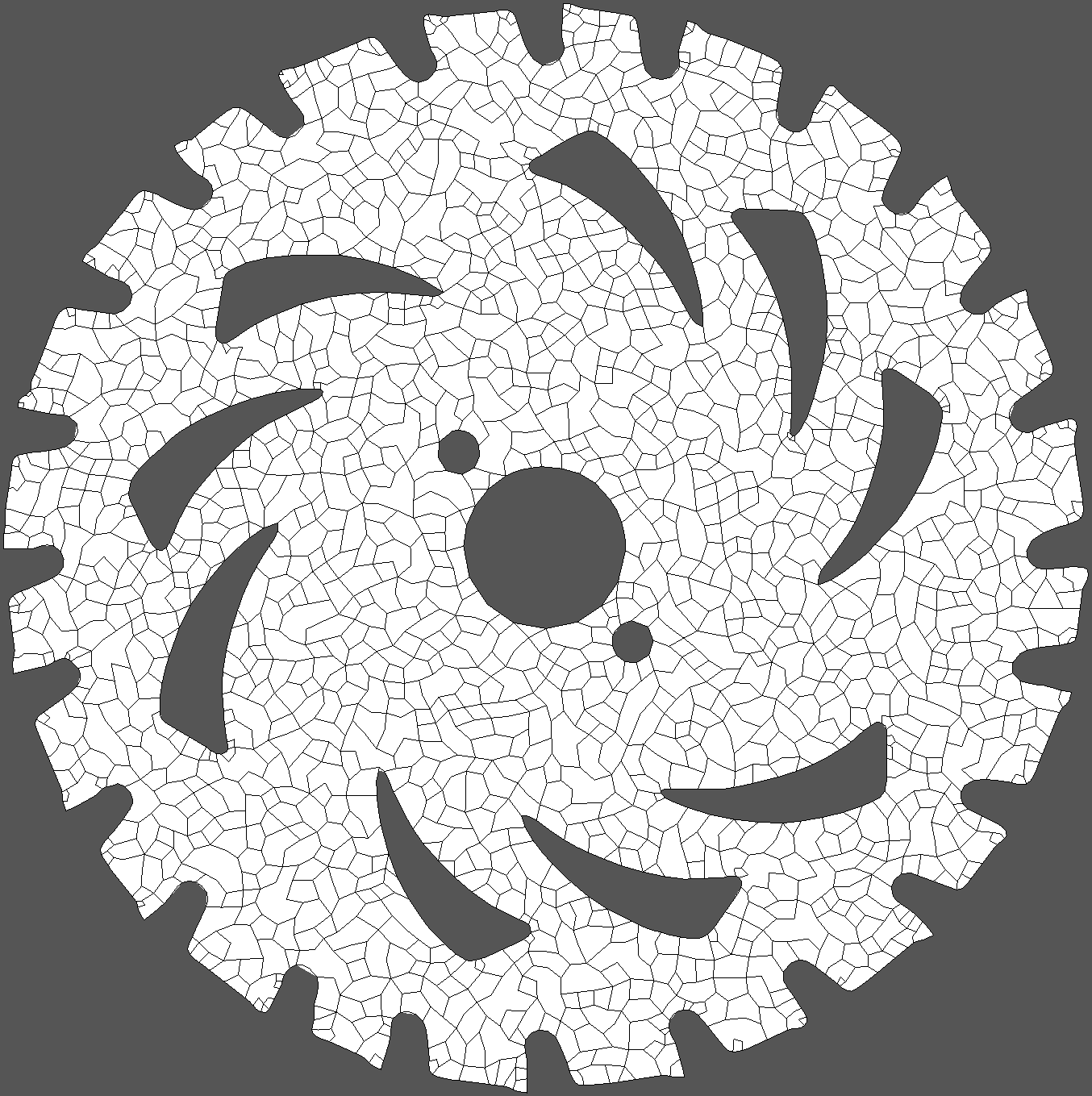}
\caption{Polygonal mesh of a circular saw generated using Polylla mesh generator.}
\label{fig:boat1}
\end{figure}

To overcome these limitations, we've introduced Polylla, an algorithm designed for generating meshes using arbitrary polygonal shapes. Polylla begins by taking a triangulation as input and then constructs terminal-edge regions by connecting triangles that share a common terminal edge. From these terminal-edge regions, the algorithm creates one or more convex or non-convex polygons whose boundaries are defined by edges that are not the longest edge of any triangle and/or input boundary/interface edges. Figure~\ref{fig:boat1} illustrates a polygonal mesh generated by Polylla.

In general, meshing algorithms can be classified into two groups \cite{owen1998survey, johnen2016indirect}: (i) direct algorithms: meshes are generated from the input geometry, and (ii) indirect algorithms: meshes are generated starting from an input mesh, typically an initial triangle mesh. Polylla falls in the indirect method category. Indirect methods have been a common approach to generate quadrilateral meshes by mixing triangles of an initial triangulation \cite{LeetreetoQuad, BlossonQuad, Merhof2007Aniso-5662}. The advantage of using indirect methods is that triangular meshes are relatively easy to generate because there are several robust and well-studied tools for generating constrained and conforming triangulations \cite{qhull, triangle2d, Detri2}.

Polylla offers several advantages over existing polygonal mesh generation algorithms. First, it can generate meshes with a wider range of polygonal shapes, including non-convex polygons. Second, Polylla can generate a polygonal mesh from any input triangulation.  Third, the Polylla tool  generates polygonal meshes faster than constrained Voronoi meshing tools~\cite{Salinas-Fernandez2022}.

It is known that GPU architectures are recommended to solve data parallel or close to data-parallel problems and CPU muti-core for task parallel problems \cite{navarro_hitschfeld-kahler_mateu_2014}. When designing a new parallel algorithm, it is important to keep such concepts in mind. We seek to answer the following research questions in this work:

\begin{itemize}
\item Can terminal-edge regions be built from triangulations using a data-parallel approach?
\item Can polygons obtained from terminal-edge regions be built using a data-parallel approach?
\item Which data structure provides an efficient time and storage performance to manage the mesh topology in parallel?
\item Is it possible to efficiently manage the creation of polygons of arbitrary size in GPU architectures?
\item What is the maximum capacity in terms of mesh size and vertices that the current GPU implementation can support? What are the limitations regarding the number of vertices in generating meshes on the GPU?
\end{itemize}

This work presents the design and implementation of GPolylla, a GPU-accelerated  Polylla that benefits from the GPU architecture by using a massive amount of threads to process in parallel each edge of the input triangulation. 
This implementation works by representing a triangulation $\tau = (V,E)$ as a half-edge data structure in GPU, and changing the values of the attributes next and prev of each half-edge to unlink edges without need of allocating/deallocating memory in GPU. 

The original Polylla algorithm does not use half-edge data structure, a new version the Polylla algorithm was presented in~\cite{salinas2023generation}, but it only uses this data structure for the input mesh. Thus, in this paper, we also present a new implementation of the Polylla algorithm that have as input and output the half-edge data structure, in order to make it comparable with the parallel version in the speed up calculation.

In Section~\ref{sec:relatedwork} we present the related work. Section~\ref{sec:polylla} describes the Polylla algorithm and the basic concepts to understand the implementations in secuental and GPU. Section~\ref{sec:datastructure} describes the half-edge data structure and its implementation in GPU. \ref{sec:sequentialpolylla} describes a new the Polylla algorithm implementation that uses the same system of removing haklf-edge by change the adjacencies of the operations next and prev. Section~\ref{sec:Gpolylla} describes the proposed parallel algorithm. Section~\ref{sec:experiments} presents the experimental results. Finally, Section~\ref{sec:conclusions} presents the conclusions and future work.

\subsection{Contribution}

The contribution we make to the scientific community through this work as:

\begin{itemize}
    \item A new version of the secuential Polylla algorithm that have as output a polygonal mesh with the half-edge data structure.
    \item An algorithm to generate polygonal meshes with arbitrary shapes accelerated by the GPU.
    \item A novel way to use the half-edge data structure in GPU for mesh generation in general.
    \item A data parallel  algorithm to build the longest-edge propagation path(Lepp) that can be useful to accelerate usign the GPU  refinement and optimizatin algorithms based on the Lepp.
    \item An example of how to use tensor core technology to accelerate phases of a meshing tool.
\end{itemize}

\section{Related Work} \label{sec:relatedwork}

The process of generating a good quality mesh usually involves three main steps \cite{Designingaproductfamilyofmeshingtools, Bern00meshgeneration}: (i) Generation of an initial mesh; (ii) Refinement of the mesh, and (iii) Optimization of the mesh. The generation of an initial mesh involves making a mesh over a given geometric domain $\Omega$. There are several methods to generate an initial mesh. Common approaches to generate unstructured meshes include Delaunay methods \cite{cheng2013delaunay}, Voronoi diagram methods \cite{YAN2013843, 2dcentroidalvoro, talischi2012polymesher}, advancing front methods \cite{AdvancingfrontmethodORIGINAL, lohner1996progress}, quadtree-based methods \cite{bommes2013quad}, and hybrid methods \cite{owen1999q}.

To enhance the efficiency of mesh generation, several parallel mesh generation algorithms have been developed. These algorithms decompose the original mesh generation problem into smaller sub-problems, which are solved in parallel using multi-threading or multi-core methods \cite{ParallelMeshGeneration, Paralleladvancingfront, ParallelDelaunay}. However, research on 2D and 3D mesh generation that leverages GPU parallelization is relatively scarce. 

A few approaches have been proposed to generate the Voronoi diagram in GPU. One approach  uses the z-buffer to generate Voronoi diagrams from images~\cite{VoronoiGPU}, and 
another  applies the Parallel Banding Algorithm(PBA) on the GPU for computing the precise Euclidean Distance Transform (EDT) of binary images in  2D and 3D, and uses both concepts to generate 2D and 3D Voronoi diagrams in GPU~\cite{VoroGPU}. A similar approach has been used to generate 
 2D Delaunay triangulations~\cite{Delaunaygpu} from a point set,   mapping the points(sites) to a texture,   computing a Voronoi diagram, generating triangles, and making necessary adjustments through a ten-step process, involving both GPU and CPU operations while ensuring consistent triangle orientations and avoiding duplicates. This algorithm was extended to work with constrained Delaunay triangulations in~\cite{6361389}. That work constructs a Constrained Delaunay Triangulation (CDT) on the GPU, combining principles from two categories of CDT construction methods. It consists of five phases: digital VD construction (using the PBA), triangulation construction, shifting, missing points insertion, and edge flipping. The two first phases use a hybrid CPU-GPU approach, and the rest of the algorithm is completely in GPU programming. 


In the case of indirect methods in GPU, any triangulation can be transformed into a Delaunay triangulation through the GPU-parallel edge-flipping algorithm, as proposed in Navarro et al in~\cite{NavarroHS13}. Navarro proposes an iterative algorithm based on the Delaunay edge-flip technique on GPU, which consists of two consecutive parallel computation phases in each iteration. The first phase is the detection of non-Delaunay edges, exclusion of edges that can not be flipped in parallel,  and processing of edges that can be flipped in parallel. The second phase is to repair the face neighborhood of the edges that were not flipped in parallel \cite{NavarroHS13}.

To our knowledge, there is no GPU-accelerated algorithm for generating polygonal meshes of arbitrary shape, and no mesh generator takes advantage of new technologies such as tensor cores in the way this research does.


%
%


\section{Polylla meshing tool} \label{sec:polylla}

This section describes main concepts and features necessaru to understand the secuential and parallel implementation of Polylla. 

The algorithm takes any initial triangulation as input $\tau = (V,E)$ to generate a polygonal mesh $\tau' = (V,E')$. The algorithm merges triangles to generate polygons of arbitrary shape (convex and non-convex shapes) according to some criterion. In Polylla we use the Longest-edge Propagation Path(Lepp)~\cite{Rivaralepp}  criterion to cluster triangles but it can be used any other criterion.

\subsection{Meshing concepts}

To understand how the algorithm works, we must introduce first the concepts of longest-edge propagation path, terminal-edge regions, terminal-edge, and frontier-edges.

\begin{definition}
\textbf{Longest-edge propagation path}~\cite{Rivaralepp} For any triangle $t_0$ of any conforming triangulation $\tau$, the Longest-Edge Propagation Path of $t_0$ ($Lepp(t_0)$) is the ordered list of all the triangles $t_0, t_1, t_2,  ..., t_{n-1}$,  such that $t_i$ is the neighbor triangle of $t_{i-1}$ by the longest edge of $t_{i-1}$, for $i = 1,2, ..., n$. The longest-edge adjacent to $t_n$ and $t_{n-1}$ is called \textbf{terminal-edge}.
\end{definition}

\begin{definition}
\textbf{Terminal-edge region} \cite{Ascom209} A terminal-edge region $R$ is a region formed by the union of all triangles $t_i$ such that Lepp($t_i$) has the same terminal-edge.
\end{definition}

\begin{definition}
\textbf{Frontier-edge} \cite{Ascom209} A frontier-edge is an edge that is shared by two triangles, each one belonging to a different terminal-edge region.
\end{definition}

Notice that each terminal-edge region $R \in \tau$ is surrounded by frontier edges.  A special case of frontier edges are barrier edges, which are frontier edges shared by two triangles belonging to the same terminal-edge region. An endpoint of a \emph{barrier edge} that belongs to only one frontier edge is called a \emph{barrier tip}.

\begin{figure}[htbp]
  \centering
\begin{subfigure}[b]{0.48\textwidth}
\includegraphics[width=\textwidth]{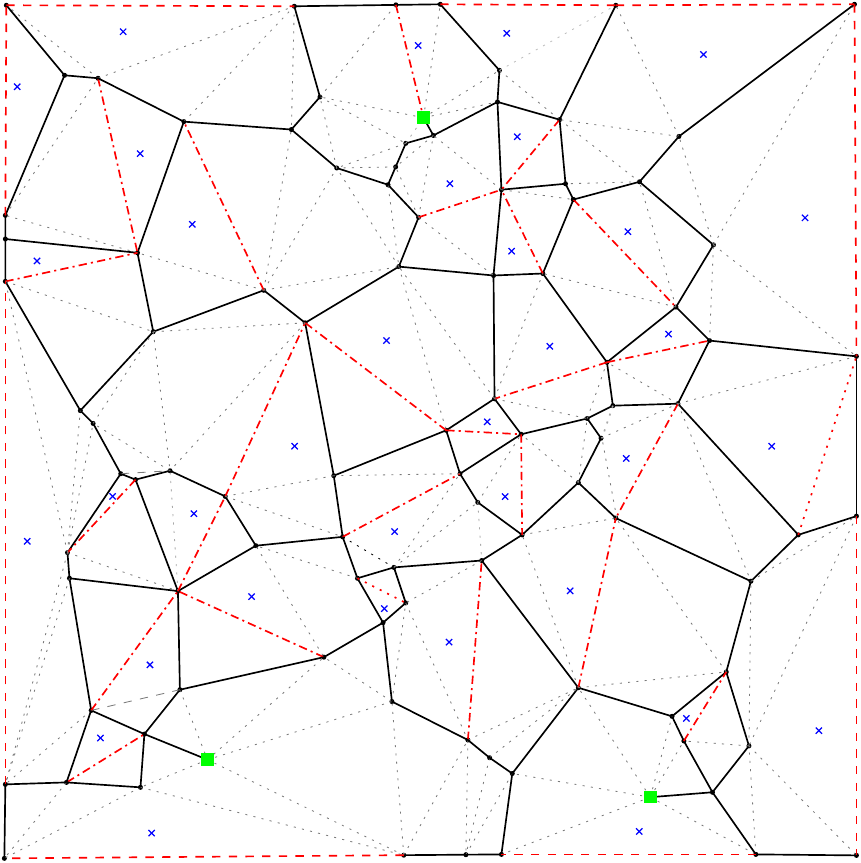}
\caption{Label Phase}
\label{fig:labelphase}
\end{subfigure}
\begin{subfigure}[b]{0.48\textwidth}
\includegraphics[width=\textwidth]{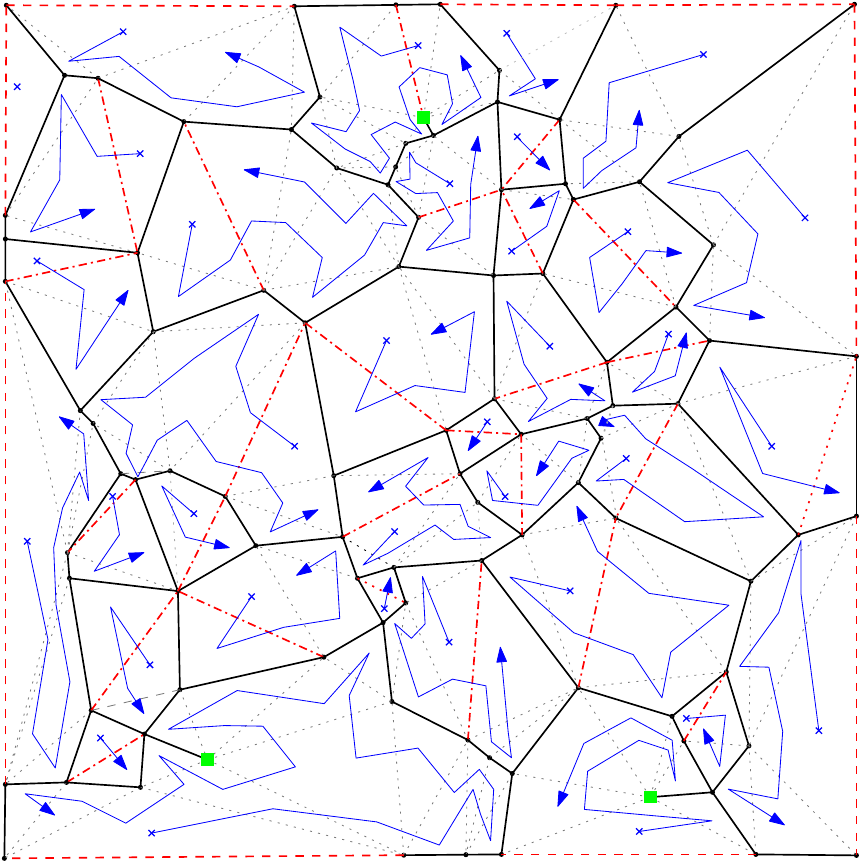}
\caption{Traversal phase}
\label{fig:traversalphase}
\end{subfigure}

\vspace{0.5cm}

\begin{subfigure}[b]{0.48\textwidth}
\includegraphics[width=\textwidth]{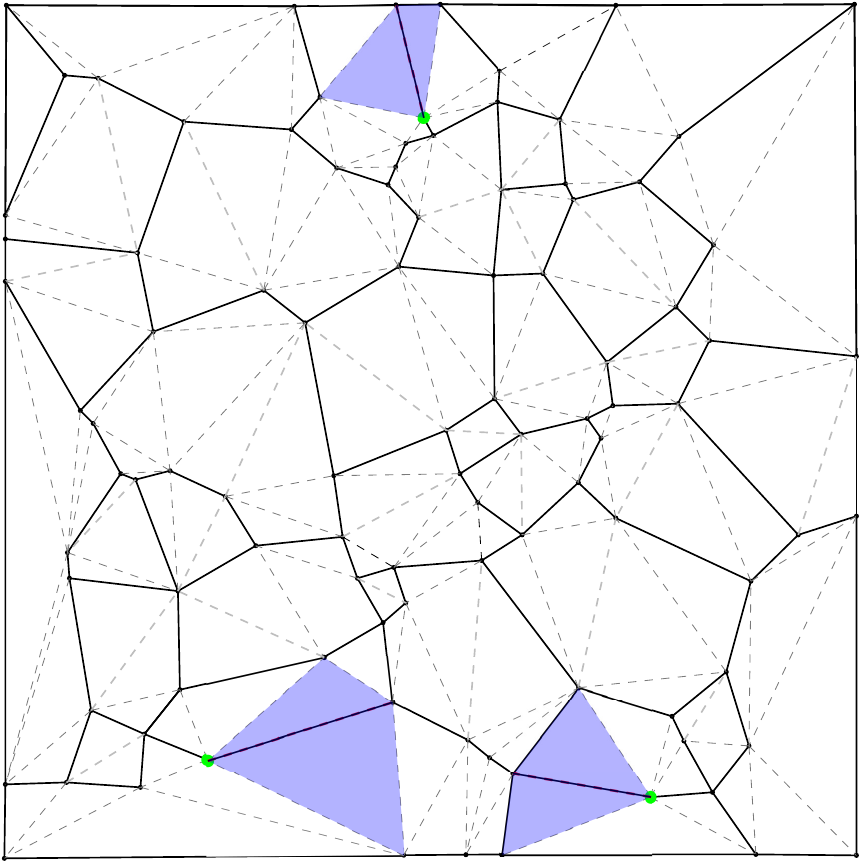}
\caption{Repair phase}
\label{fig:repairphase}
\end{subfigure}
\begin{subfigure}[b]{0.48\textwidth}
\includegraphics[width=\textwidth]{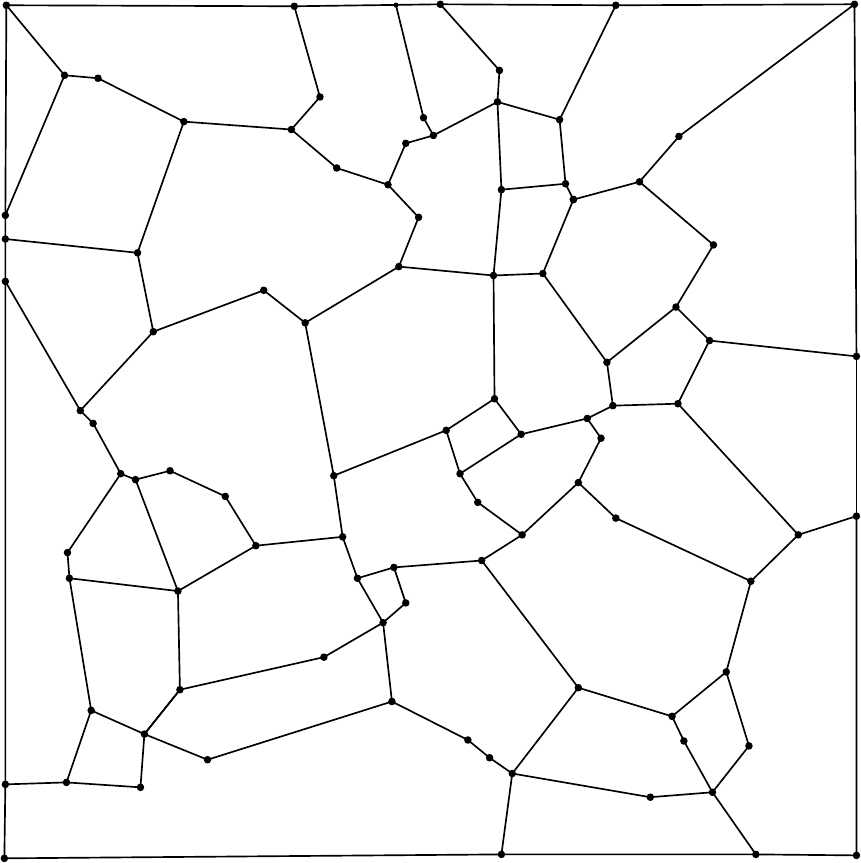}
\caption{Final mesh}
\label{fig:finalmesh}
\end{subfigure}
  \caption{Example of generating a Polylla mesh from an initial triangulation: (a) Output of the label phase to generate terminal-edge regions. Black lines are frontier edges, dotted gray lines are internal edges, and terminal edges are red dashed lines. Terminal edges can be inside or at the boundary of the geometric domain, so dashed lines are border terminal edges, and dotted dashed lines are internal terminal edges. barrier tips are green squared vertices and seed triangles are marked with a blue cross. (b) Traversal phase example: Arrows inside terminal regions show the paths of the algorithm during the conversion from a terminal-edge region to a polygon. The path starts at a triangle labeled as a seed triangle. Each terminal-edge region has only one seed triangle. (c) Example of a non-simple polygon split using interior edges with barrier tips as endpoints. To split the polygon, the algorithm labels the middle interior edges incident to the barrier tips as frontier edges, which are then stored along with cross-labeled triangles as seed triangles. The algorithm repeats the traversal phase using a new seed triangle but avoids generating the same polygon again. (d) Final Polylla mesh. Source~\cite{salinas2023generation}}
  \label{fig:Polyllaphases}
\end{figure}

\subsection{Polylla algorithm}

The Polylla algorithm performs $3$ phases to convert terminal-edge regions $R_i \in \tau$ into polygons $P\in \tau'$, those phases are show in Figure~\ref{fig:Polyllaphases} and are:

\begin{enumerate}
    
\item Label Phase: each edge in the input triangulation is labeled as a terminal edge or frontier edge based on its length and adjacency to triangles. If an edge is labeled as a terminal edge, one of its adjacent triangles is chosen as a seed triangle for the next phase.

\item Traversal Phase: one seed triangle is selected from each terminal-edge region, and each polygon is generated by traversing the vertices of the frontier edges in a counter-clockwise order. Non-simple polygons with barrier edges may be generated during this phase, which are processed in the third phase.

\item Repair Phase: non-simple polygons with barrier edges are partitioned into simple polygons. Interior edges with a barrier tip as an endpoint are used to split the non-simple polygon into two new polygons. For each new polygon, a triangle is labeled as a seed and the Traversal phase is applied to generate simple polygons. The output of the algorithm is a polygonal mesh composed of simple polygons.

\end{enumerate}

\section{Mesh representation: CPU and GPU}\label{sec:datastructure}

Choosing the right representation is crucial to achieving good computational performance in mesh processing. In the literature, various methods exist to represent mesh geometry and topology. For example, a common approach is the triangle-based mesh representation \cite{MeshDatastructSurvey}, where each triangle is represented as a set of vertices. Another option is the star-vertex representation \cite{kallmann_star-vertices_2001}, which stores important information about the mesh in the vertices. Additionally, a classic representation of a planar graph is the storage of a triangulation as an adjacency list matrix.

These three implementations are not suitable for efficient mesh generation on GPUs because they can not be easily modified to create new polygonal meshes, and because it is not possible to allocate new memory on GPUs at runtime without incurring a significant time cost. We have decided to use the half-edge data structure to represent the input and output mesh of our mesh generator because it allows us to avoid the previous issues and in addition, it is adequate for handling general polygons. The half-edge data structure, also known as Doubly Connected Edge List (DCEL)~\cite{MULLER1978217}, is an edge-based data structure that represents each edge of a polygonal mesh as two half-edges of opposite orientation. 

As we mentioned above, for polygonal meshes, the half-edge data structure is widely recognized for its flexibility and efficiency, with most essential queries having O(1) time complexity. The half-edge data structure also allows for a simple way to traverse inside a mesh using an edge or a face as a starting point. We have chosen this data structure for the sequential implementation and now for the GPU implementation as well. This data structure allows  for a natural way to parallelize the Polylla algorithm in terms of threads per edge and triangles.

For this research, we use the half-edge implementation of our previous work shown in~\cite{salinas2023generation}. The implementation contains the basic half-edge queries defined in~\cite{deBerg2008}, and some extended queries defined by us in~\cite{salinas2023generation} to facilitate the navigation inside a mesh during its generation.




\subsection{Half-edge representation}

\begin{figure}
\centering
\begin{subfigure}{0.3\textwidth}
\centering
\includegraphics[width=0.85\textwidth]{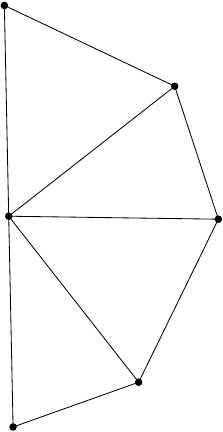}
\caption{}
\end{subfigure}
\begin{subfigure}{0.3\textwidth}
\centering
\includegraphics[width=0.85\textwidth]{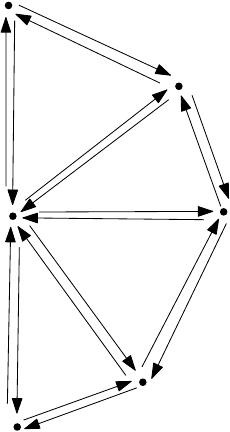}
\caption{}
\end{subfigure}
\quad
\begin{subfigure}{0.3\textwidth}
\centering
\includegraphics[width=\textwidth]{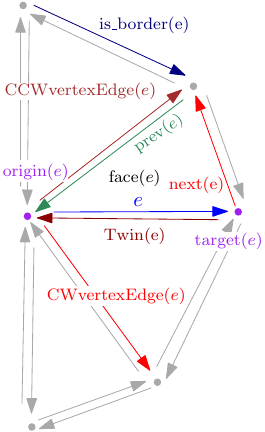}
\caption{}
\end{subfigure}
\caption{(a) Polygonal mesh (b) Representation of the mesh using half-edge data structure (b) Example of queries for a half-edge $e$. }
\label{fig:halfedge}
\end{figure}

Given a half-edge $e$ in a triangulation $\tau = (V,E)$ (see Figure~\ref{fig:halfedge}), the half-edge data structure allows traversal of the face incident to $e$ through queries such as \texttt{next($e$)} and \texttt{prev($e$)}, while \texttt{twin($e$)}  enables navigation between faces. Furthermore, the structure aims the exploration of the vertices surrounding $e$ through queries like \texttt{origin($e$)}, \texttt{target($e$)},

This data structure enables to define additional queries~\cite{salinas2023generation}. The queries \texttt{CCWvertexEdge($e$)} and \texttt{CWvertexEdge($e$)} allows traveling around the faces of $\tau$ in counterclockwise and clockwise, respectively. \texttt{is\_border($e$)} ascertains whether a half-edge is incident to the exterior face, \texttt{degree($v$)} to determine the number of edges incident to a vertex $v$, \texttt{incidentHalfedge($f$)} to obtain a half-edge incident to a face $f \in \tau$, and \texttt{edgeOfVertex($v$)} to retrieve a half-edge with origin at vertex $v$.

\begin{figure}[ht]
  \begin{minipage}{0.45\textwidth}
  \begin{lstlisting}[language=C++, caption={Vertex record}, label=listing:vertexstruct]
  struct vertex{
      double x, y;
      bool is_border;
      int incident_halfedge;
  };
  \end{lstlisting}
  \end{minipage}
  \hfill
  \begin{minipage}{0.45\textwidth}
  \begin{lstlisting}[language=C++, caption={Half-edge record}, label=listing:halfedgestruct]
  struct halfEdge {
      int origin;
      int twin; 
      int next, prev;
      bool is_border;
  };
  \end{lstlisting}
  \end{minipage}
  \caption{Vertex and Half-edge records}
  \end{figure}

\subsection{Half-edge: Cuda implementation}

The CUDA implementation of this data structure can be achieved using two Arrays of Structures (AoS), namely the \texttt{Halfedge array} and the \texttt{Vertex array}. These arrays provide access to the mesh information. For a detailed view of the implementation, see the Listing~\ref{listing:vertexstruct} and \ref{listing:halfedgestruct}.


The half-edge data structure stores some information implicitly, which enhances its efficiency. For a triangulation, each three half-edges in the \texttt{Halfedge array} represent a face, enabling the \texttt{incidentHalfedge($f$)} query through the formula $3\cdot \#faces$.  \texttt{CCWvertexEdge($e$)} and \texttt{CWvertexEdge($e$)} queries can be implemented using \texttt{twin(next($e$))} and \texttt{twin(prev($e$))}, respectively. The query target is defined as \texttt{twin(origin($e$))}. The \texttt{degree($v$)} query is achieved by cycling around an edge incident to $v$ using the \texttt{CCWvertexEdge($e$)} query. 

\begin{figure}[h!]
  \centering
  \begin{subfigure}[b]{0.3\linewidth}
    \includegraphics[width=\linewidth]{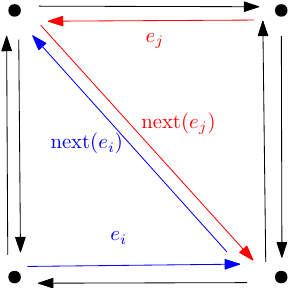}
    \caption{}
    \label{fig:join_faces}
  \end{subfigure}
  \hfill
  \begin{subfigure}[b]{0.3\linewidth}
    \includegraphics[width=\linewidth]{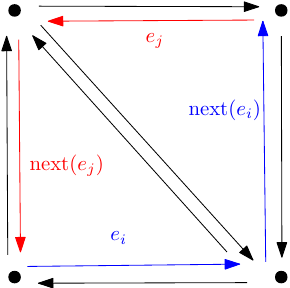}
    \caption{}
    \label{fig:join_faces2}
  \end{subfigure}
  \hfill
  \begin{subfigure}[b]{0.3\linewidth}
    \includegraphics[width=\linewidth]{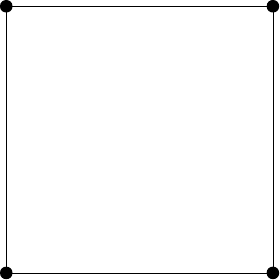}
    \caption{}
    \label{fig:join_faces3}
  \end{subfigure}
  \caption{\textbf{(a)} A terminal-edge region formed by two adjacent triangles represented as the half-edge data structure. \textbf{(b)} The terminal-edge is  dereferenced  by changing the values of the next and prev of $he_i$ and $he_j$. \textbf{(c)} Final result of the edge elimination, a square. }
  \label{fig:join_faces_all}
\end{figure}

The half-edge data structure is first built on the CPU, then the \texttt{HalfEdge array} and \texttt{Vertex array} are copied to the device. On the device, the value of \texttt{next} and \texttt{prev} attributes of each half-edge are changed in such way that the input is converted from a triangular mesh to a Polylla mesh.  
In order to explain this strategy, Figure~\ref{fig:join_faces_all} shows two triangles being joined to create a square. In Figure~\ref{fig:join_faces} a polygonal mesh with two faces represented the half-edge data structure. On the other hand, in Figure~\ref{fig:join_faces2} a change of the attribute values in the half-edges $he_i$ and $he_j$ to \texttt{next($he_i$)} = \texttt{CWvertexEdge(next($he_i$))} and \texttt{next($he_j$)} = \texttt{CWvertexEdge(next($he_j$))} is shown. Finally, Figure~\ref{fig:join_faces3} shows the resulting polygon from joining the two triangles' faces. 
The same strategy applied to join triangles can be extended to join polygons. This strategy is explained with more details in Section~\ref{sec:CPUtraversalphase}.

This method allows us to work in the GPU without involving edge removal, which means we do not need to change the size of the allocated initial memory. The algorithm just unlinks the half edges not needed anymore. If we want to change the size of the allocated memory, we would need to copy the half-edge AoS back to the host, call the CUDA free function, ask for  the new memory with CUDA malloc, and copy the new half-edge AoS to the device. However, this operation is expensive, in the experiment we will see that the copy operation between host and device has a high cost.

It is worth mentioning, that keeping the half edges of the initial triangulation, allows us to implement  future mesh optimizations, such as converting non-convex polygons into convex polygons in an efficient way.

\subsection{Additional data structures}

 Prior to the parallel implementation of the Polylla algorithm utilizing the half-edge data structure, it is necessary to establish several additional temporary data structures. Those extra data structures are different between CPU and GPU.

\textbf{Secuential}: In order to label each edge of the triangulation, two bit-vectors, namely {\tt longest-edge bitvector} and {\tt frontier-edge bitvector}, are utilized to indicate the longest edge of each triangle and frontier edges, respectively. A bit set to 1 means that the corresponding half-edge is a longest-edge or frontier-edge, respectively. The length of both bit-vectors is $2\vert E \vert$, where $\vert E \vert$ represents the number of edges of the triangulation. For the seed triangles, a  {\tt seed-list} stores the indices of the incident terminal edges.

In the traversal phase we do a copy of the \texttt{half-edge array} to change the values of the attributes \textsc{next} and \textsc{prev}, and with this, represent the half-edge of the new polylla mesh $\tau'$. Despite this copy is optional in the secuential version, we decide to this to match with the GPU version.

During the Repair phase, two auxiliary arrays are employed to prevent the duplication of polygons, namely, an initially empty \texttt{subseed-list} and a \texttt{usage bitvector} of length $|E|$.

Finally, after generate the Polylla mesh, the seed list is used to rebuild each polygon of the output mesh using the next and previous queries on that half-edge.


\textbf{GPU}: The GPU implementation uses the same data structure, the  arrays {\tt longest-edge bitvector} and {\tt frontier-edge bitvector} are also the same, in GPU the  {\tt seed-list} is a  {\tt seed-bitvector} of size $2\vert E \vert$, and after generate the Polylla mesh the bitvector is converted to an {\tt output seed edges} of integers to facilitate the process of access and traverse in the polygons of the Polylla mesh.

\section{Secuential Polylla} \label{sec:sequentialpolylla} 

In this section we will talk about a new version of Polylla, used to compare the GPU accelerated version. 

The version presented in~\cite{Salinas-Fernandez2022} used a triangle data structure to generate Polylla meshes. The version showed in~\cite{salinas2023generation} uses a half-edge version as input but as output have a face based data structured. And the new version presented in this paper have as input and output the half-edge data structure, this have the advantage that we can use the half-edge queries in the Polylla meshes, and we can use the same data structure for Secuential and GPU implementation to compare both versions.

This new version have the same 3 phases, the label phase, the traversal phase and the repair phase. The only version in comparison~\cite{salinas2023generation} that change is the traversal as instead of store the polygons in a face based data structure, we copy half-edge data structure to the output mesh and modify the decencies of the queries \texttt{next} and \texttt{prev} to unlink the edges that are not part of the output mesh.

The figures useful to understand each phase are the showed in Figures~\ref{fig:Polyllaphases}, so we will not repeat them here.

\subsection{Label phase}

This phase receives a triangulation $\tau = (V,E)$ as input. The objective is to label frontier-edges to identify the boundary of each terminal-edge region $R_i$ in $\tau$ and to select one triangle of each $R_i$ to be used as the seed for generating new polygons in the next phase, the Traversal phase. To do this, the phase first finds the longest edge of each triangle in $\tau$, and then labels the frontier edges and the seed edges. 

An example of a resulting labeled triangulation after this phase is shown in Figure~\ref{fig:labelphase}.

\begin{algorithm}
  \caption{Secuential Polylla main algorithm}\label{algo:Polylla algorithm}
  \begin{algorithmic}[0]
  \Require Labeled triangulation $\tau_L = (V,E)$
  \Ensure Polylla mesh 
  \State Label phase \Comment{Algorithm \ref{algo:CPU_labelmaxedge}, \ref{algo:CPUlabelfrontieredge}, \ref{algo:CPUlabelfseeddedge}}
  \State Traversal phase \Comment{Algorithm \ref{algo:CPUtravelphase}}
  \State Repair Phase \Comment{Algorithm \ref{algo:CPURepairphase}}
  \end{algorithmic}
\end{algorithm}

\subsubsection*{Label longest-edge}\label{sec:CPUlongesedge}

For each triangle $t_i \in \tau$, composed by 3 half-edges, calculates which half-edge is the longest. 

To calculate the longest edge of each triangle $t_i \in \tau$ the algorithm iterates sequentially over each triangle $t \in \tau$,  obtains the half-edge incident to $t_i$, calculates the length of the half-edges $he_i$, $next(e_i)$, $prev(e_i)$, and stores longest edge information in the  \texttt{longest\_edge\_bitvector} as shown in algorithm \ref{algo:CPU_labelmaxedge}.

\begin{algorithm}
\footnotesize

    \caption{Secuential Label phase: Longest edge labeling}\label{algo:CPU_labelmaxedge}
    
    \begin{algorithmic}[0]
    \Require Triangulation $\tau = (V,E)$
    \Ensure Labeled triangulation $\tau_L = (V,E)$
    \Function{CPU  longest edge labeling}{$\tau$, \texttt{longest\_edge\_bitvector}}
    \ForAll{triangle $t_i$ in  {\tt HalfEdge}} \label{algolabel:TriangleIteration}
        \State $he_i \leftarrow$ incident half-edge to $t_i$
        \State $d_1, d_2, d_3 \leftarrow$ length size of $he_i$, $next(he_i)$, $prev(he_i)$
        \State $he_{max} \leftarrow$ $max(d_1, d_2, d_3)$ 
        \State \texttt{longest\_edge\_bitvector}[$he_{max}$] = True 
    \EndFor
    \EndFunction

    \end{algorithmic}
\end{algorithm}

\subsubsection*{Label frontier-edges}\label{sec:CPUlabelfrontieredge}

This algorithm is shown in Algorithm~\ref{algo:CPUlabelfrontieredge}. The algorithm  computes whether $he_i$ is a frontier-edge for each half-edge $he_i \in \tau$. This step takes place after the longest edge of each triangle was found and labeled in the previous phase. Thus, the algorithm for each half-edge $he_i \in \tau$ asks if $he_i$ or \texttt{twin($he_i$)} are not labeled as the longest-edge of its incident triangle in \texttt{longest\_edge}, and if $he_i$ or \texttt{twin($he_i$)} are border half-edge, if one of both question is true, it is sorted in {\tt frontier-edge}[$he_i$] as true, this mean, it labeled $he_i$ as a frontier-edge.

\begin{algorithm}
\footnotesize

    \caption{Secuential Label phase: Label frontier edges}\label{algo:CPUlabelfrontieredge}
    \begin{algorithmic}[0]
    \Require Triangulation $\tau = (V,E)$
    \Ensure Labeled triangulation $\tau_L = (V,E)$
    \BKernel{\texttt{LabelFrontierEdges}($\tau = (V,E))$}
    \ForAll{halfedge $he_i \in \tau$ \textbf{in parallel}} 
        \State \texttt{is\_not\_longest\_edge?} $\leftarrow$ $he_i$ and twin($he_i$) are not longest\_edges
        \State \texttt{is\_border\_edge?} $\leftarrow$ $he_i$ or twin($he_i$) is a boundary edge
        \If{\texttt{is\_longest\_edge?} or is\_border\_edge?}
            \State Label $he_i$ as frontier-edge
        \EndIf    
    \EndFor
    \EKernel
    \end{algorithmic}
\end{algorithm}

\subsubsection*{Label seed-edges}

This algorithm is shown in Algorithm~\ref{algo:CPUlabelfseeddedge}. In this step, we select a half-edge inside a terminal-edge region as a seed-edge, to be used to create a new polygon in the Traversal Phase. The algorithm iterates over each half-edge $he_i \in \tau$, for each half-edge the algorithm checks if $he_i$ is a terminal edge or a terminal-border edge, this is done by calculating if $he_i$ and twin($he_i$) are labels as the longest-edge in \texttt{max\_edge}, if it is the case, one of both $he_i$ or twin($he_i$) is store in a list of integers called {\tt seed-list} to use it in the Traversal phase to generate a new polygon.

\begin{algorithm}
  \footnotesize
  
      \caption{Secuential Label phase: Label seed edges}\label{algo:CPUlabelfseeddedge}
      \begin{algorithmic}[0]
      \Require Triangulation $\tau = (V,E)$
      \Ensure Labeled triangulation $\tau_L = (V,E)$
      \Function{\texttt{LabelSeedEdges}}{$\tau = (V,E)$}
      \ForAll{halfedge $he_i \in \tau$} 
          \State \texttt{is\_terminal\_edge?} $\leftarrow$ $he_i$ and twin($he_i$) are max edges and not border 
          \State \texttt{is\_terminal\_border\_edge?}  $\leftarrow$ $he_i$ or twin($he_i$) is max edge and border
          \If { \texttt{is\_terminal\_edge?} or \texttt{is\_terminal\_border\_edge?} }
              \State Label $he_i$ or twin($he_i$) as seed-edge
          \EndIf    
      \EndFor
      \EndFunction
      \end{algorithmic}
  \end{algorithm}

  \subsection{Traversal Phase}\label{sec:CPUtraversalphase}

\begin{figure}[h]
    \centering
    \includegraphics[width=0.5\linewidth]{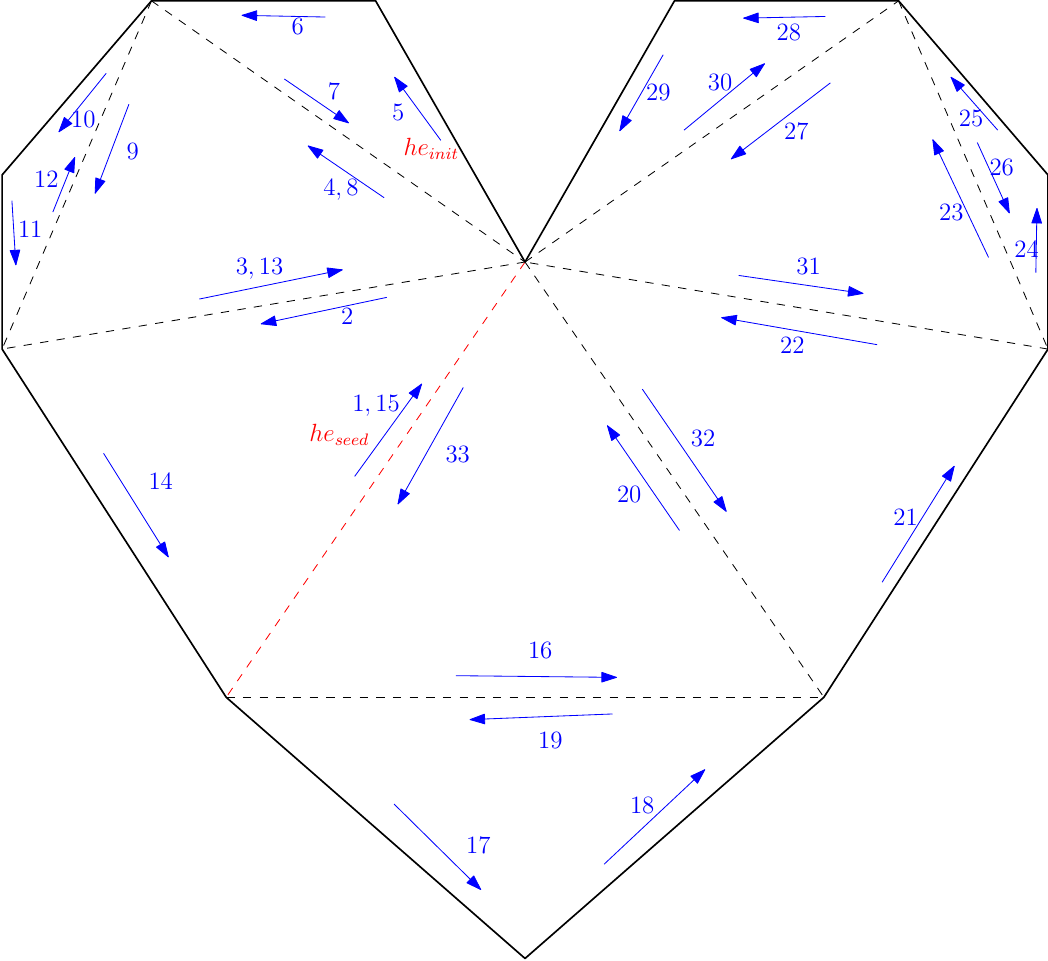}
    \caption{Terminal-edge region $R_i$ to convert into a new polygon. The seed edge of $R_i$ is the half-edge $he_{seed}$, which is the terminal-edge of $R_i$. As the algorithm needs to traverse inside $R_i$ using the frontier-edges of $R_i$, the algorithms seach an frontier-edge near to $he_{seed}$ to define a $he_{init}$ and do the traversal seaching for the next frontier-edge continous to $he_{init}$. Blue arrows are the half-edges and the number is order in which each half-edge is visited by the algorithm.}
    \label{fig:cpu_travel_corazon}
  \end{figure}

\begin{figure}[h]
    \centering
    \includegraphics[width=0.5\linewidth]{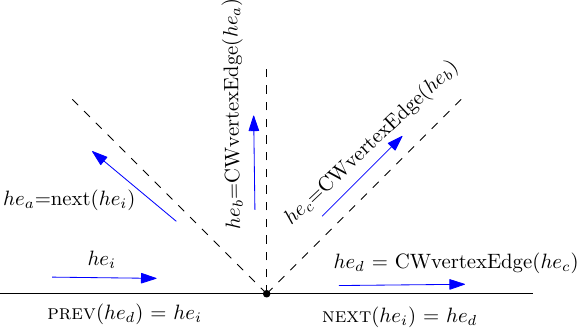}
    \caption{Example of how the algorithm uses the function \textsc{CWvertexEdge} to circle around a vertex and change the value of the next of a half-edge $he_i$. In the Figure the lines are frontier-edges, dashed lines are internal-edges and arrows are half-edges. The algorithms wants to change the \textsc{next} attribute of $he_i$, thus this circle around \textsc{target}($he_i$) until find another frontier-edge $he_d$ and defines \textsc{next}($he_i$) = $he_d$ and \textsc{prev}($he_d$) =  $he_i$}.
    \label{fig:cpu_travel}
  \end{figure}

  In this phase we traversal inside each terminal-edge region $R_i \in \tau$ to generate a polygon $P \in \tau'$, this phase is shown in Figure~\ref{fig:traversalphase}.  The objective of this traversal is traveling inside the frontier-edges of a terminal-edge region $R_i$, see Figure~\ref{fig:cpu_travel_corazon}, and change the adjacencies of two continuous frontier-edges as in show in Figure~\ref{fig:cpu_travel}. By changing the values of the attributies \textsc{next} and \textsc{prev}, the algorithm generate each polygon of $\tau'$ in an implict way as was showed in Figure~\ref{fig:join_faces_all}.

  As it was explained in section \ref{sec:datastructure}, the triangulation $\tau = (V,E)$ is represented as a half-edge data structure in a array called \texttt{mesh\_input array}. In this phase we copy the \texttt{mesh\_input array} to a \texttt{mesh\_output array} to store the output mesh $\tau' = (V,E')$ as a half-edge data structure. The \texttt{mesh\_output array} is the half-edge data structure that will change the values of the \texttt{next} and \texttt{prev} attributes to generate the polylla mesh $\tau'$. 

  The traversal phase is shown in Algorithm~\ref{algo:CPUtravelphase}. This algorithm is a function that receives a seed edge $he_i$, if $he_i$ is a internal-edge, the algorithm circles around \textsc{target}($he_i$) until find a frontier-edge to use as begining of the traversal. Now, the algorithms defines the indices $in\,curr$, defined as the index of \textsc{next}($he_{init}$), and $out\,curr$, defined as the index $init$. Afterward define the initial conditions, the algorithm traverse around the terminal-edge region $R_i$ repeating the operation show in Figure~\ref{fig:traversalphase} for each frontier-edge in $R_i$ until $in\,curr = init$. 

  Afterward, the algorithm set the frontier-edge $he_{init}$ a seed thus the algorithm can use it to generate the polygon of mesh again.

  Notice that not all the polygons are useful for the context of meshes, some of them are non-simple polygons with barrier tips, example of those poylgons are showed in~\ref{fig:cpu_travel_corazon}, the barrier tips are the green vertices. To check if a generated polygon contain barrier-edges, we start in $he_{init}$ and we traverse $P$ using the query \texttt{next($he_i$)}, for each visited edge $he_i$ we check if $he_i ==$ \texttt{twin($he_i$)}, if this is the case, then $he_i$ is a barrier-edge, the polygon $P$ needs to be repaired in next phase, the repair phase. 

  \begin{algorithm}
  \footnotesize
  
\caption{Secuential Traversal phase: Polygon construction}\label{algo:CPUtravelphase}
\begin{algorithmic}[1]
      \Require Seed edge $he$, Input Half-edge array \texttt{mesh\_input}, Output Half-edge array \texttt{mesh\_output} 
      \Ensure Output Half-edge array \texttt{mesh\_output} with the half-edge attributes modified.
      
      \While {$he$ is not a frontier-edge} 
          \State $he$ $\leftarrow$ CWvertexEdge($he$)
      \EndWhile

      \State ${init}$ $\leftarrow$ index of $he$
      \State ${in\,curr}$ $\leftarrow$ \textsc{next(\texttt{mesh\_input[}$init$\texttt{]})}

      \State ${out\,curr}$ $\leftarrow$ ${init}$ 

      \Repeat
          \While {\texttt{mesh\_input[}$init$\texttt{]} is not a frontier-edge} \label{algotravel:end}
              \State ${in\,curr}$ $\leftarrow$ \textsc{CWvertexEdge}(\texttt{mesh\_input[}$init$\texttt{]})
          \EndWhile

          \State \texttt{mesh\_output[}$out\,curr$\texttt{]}.next $\leftarrow$ $in\,curr$
          \State \texttt{mesh\_output[}$in\,curr$\texttt{]}.prev $\leftarrow$ $out\,curr$

          \State $in\,curr$ $\leftarrow$ \textsc{next}(\texttt{mesh\_input[}$in\,curr$\texttt{]})
      \Until{${init} = in\,curr$}
      \State Set half-edge \texttt{mesh\_output[}$init$\texttt{]} as seed edge
\end{algorithmic}
\end{algorithm}

\subsection{Repair phase} \label{sec:CPURepairPhase}

This phase is shown in the algorithm~\ref{algo:CPURepairphase}. This algorithm is a polygon $P$ with barrier tips, and it splits the polygon until it removes all barrier tips. The output is a set of seed half-edges that represent the new polygons generated in this phase. Despite on the changes made in the label phase and traversal phase, this phase is the same as the showed in~\cite{Salinas-Fernandez2022,salinas2023generation} without any modification.

\begin{figure}[h]
  \centering  
  \resizebox{0.8\linewidth}{!}{
  \subfloat[] {\label{fig:betosplit}\includegraphics[width=0.3\textwidth, ]{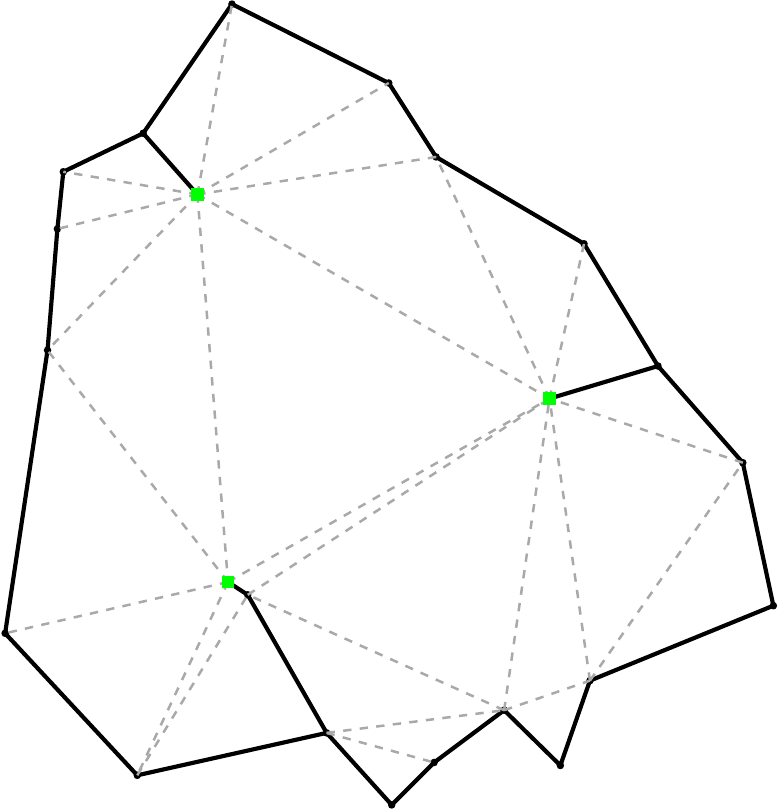}}\hspace{0.5cm}
  \subfloat[]{\label{fig:besplited}\includegraphics[width=0.3\textwidth]{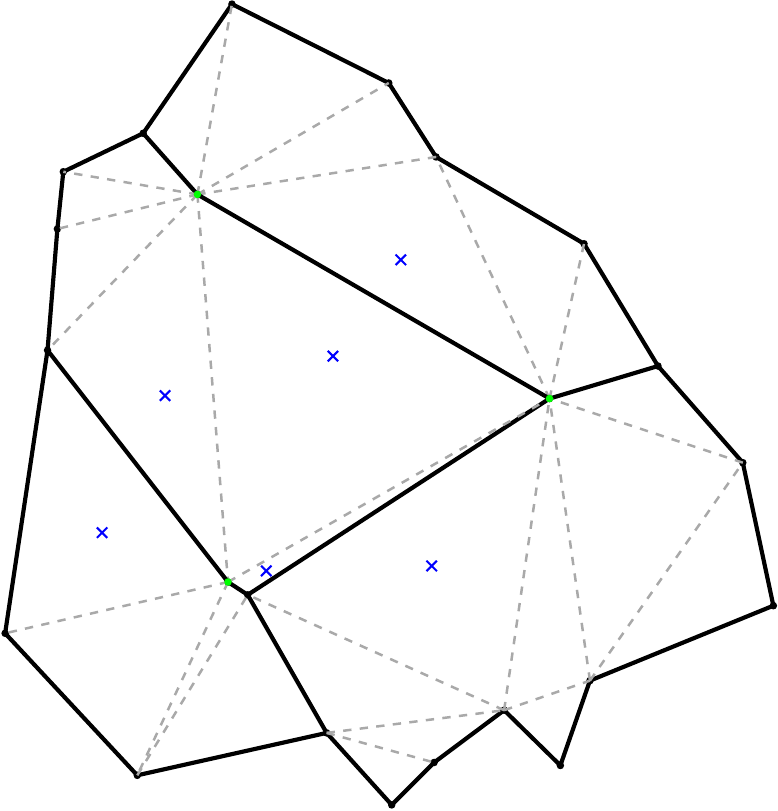}}\hspace{0.5cm}
  \subfloat[]{\label{fig:benewpolygons}\includegraphics[width=0.3\textwidth]{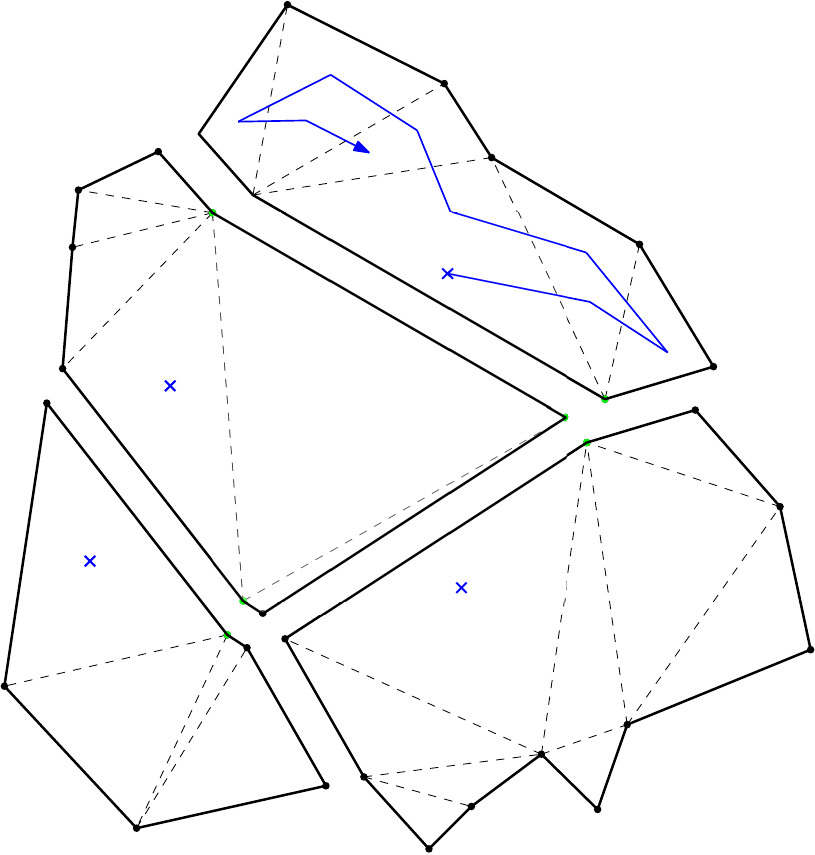}}
  }
  \caption{Example of a non-simple polygon split using interior edges with barrier tips as endpoints. (a) Non-simple polygon. (b) Middle interior edges incident to barrier tips are labeled as frontier-edges (solid lines), and cross-labelled triangles are stored as seed triangles. (c) The algorithm repeats the travel phase using a new seed triangle but avoiding generating the same polygon again. Source:~\cite{Salinas-Fernandez2022,salinas2023generation}.}
  \label{figs:splitmid} 
  \end{figure}

For each barrier tip $b_i \in P$, it  calculates the middle edge that contains $b_i$. It then labels the edges of the middle edge as frontier-edges, uses them as seed edges, stores in a list them in a list called \texttt{subseed-list}, and labels them as {\tt True} in \texttt{usage bitvector}, to mark them as visited during this phase.

After finding all barrier tips, the algorithm generates polygons from the half-edge $he_i \in$ \texttt{subseed-list} using the Algorithm~\ref{algo:CPUtravelphase} of the traversal phase, only half-edge that are in \texttt{subseed-list} and \texttt{usage bitvector} are used as seed-edges. If a half-edge of the \texttt{subseed-list} is visited during the traversal, then it is removed from \texttt{subseed-list} and labeled as \texttt{False} in \texttt{usage bitvector}.

\begin{algorithm}
\footnotesize
\caption{Secuential Repair phase}\label{algo:CPURepairphase}
\begin{algorithmic}[0]
\Require Seed edge of a non-simple polygon $P$
\Ensure Set of simple polygons $S$
\State Initialize \texttt{subseed list} as $L_p$ and \texttt{usage bitarray} as $A$ \label{algorepa:inithash}
\State $S \leftarrow \emptyset$
\ForAll{barrier tip $b$ in $P$} \label{algorepa:foreachbet}
\State $e \leftarrow$ \textsc{edgeOfVertex($b$)} \label{line:searchfrontier}
\While{$e$ is not a frontier-edge}
\State $e \leftarrow$ \textsc{CWvertexEdge($e$)}
\EndWhile\label{line:searchfrontier2}
\For{0 to $(\textsc{degree}(b)-1)/2$} \label{line:midedge}
\State $e \leftarrow$ \textsc{CWvertexEdge($e$)}
\EndFor \label{line:endmiodege}
\State Label $e$ as frontier-edge \label{line:labelmidedge}
\State Save half-edges $he_1$ and $he_2$ of $e$ in $L_p$ \label{line:savebothhalfedges}
\State $A[he_1] \leftarrow \texttt{True}$, $A[he_2] \leftarrow \texttt{True}$ \label{line:savetobivector}
\EndFor
\ForAll{half-edge $h$ in $L_p$} \label{algorepa:foreachseedtriangle}
\If{$A[h]$ is \texttt{True}} \label{algorepa:true}
\State $A[h] \leftarrow \texttt{False}$
\State Generate new polygon $P'$ starting from $h$ \label{algorepa:generationpoly} using Algorithm \ref{algo:CPUtravelphase}. 
\State Set as \texttt{False} all indices of half-edges in $A$ used to generate $P'$ \label{algorepa:removeseeds}
\EndIf

\State $S \leftarrow S \cup P'$ 
\EndFor 
\State \Return $S$
\end{algorithmic}
\end{algorithm}

\section{GPU Polylla} \label{sec:Gpolylla}

In this section, we  introduce the GPU accelerated algorithms. The GPU variant does not have the same phases as the sequential version. In this version the label phase is almost the same as the Secuential version, but the traversal phase  and the repair phase are different. For clariffication, a kernel a function that gets executed on GPU. 

A summary of the GPU algorithm is shown in Algorithm~\ref{algo:GPU Polylla algorithm}. The GPU variant have a total of 6 kernels. In each subsection we will explain one of them.  This algorithm takes as input a  triangulation $\tau(V,E)$ and generates as output a polygonal mesh, where the output  is a half-edge representation of the polygonal mesh $\tau'(V,E)$.

In the following subsections we are going to explain each kernel called during the algorithm execution. Each kernel is called in the order showed here and with only a sync barrier between each kernel, this barrier is to force to the algorithm to wait until all threads ends to launch the next kernel.

\begin{algorithm}
  \caption{GPolylla main algorithm}\label{algo:GPU Polylla algorithm}
  \begin{algorithmic}[0]
  \Require triangulation $\tau = (V,E)$
  \Ensure Polylla mesh $\tau' = (V,E)$
    \State Label the longest-edge of each triangle \Comment{Algorithm \ref{algo:gpu_labelmaxedge}}
    \State Label  frontier edges \Comment{Algorithm~\ref{algo:GPUlabelfrontieredge}}
    \State Label seed edges \Comment{Algorithm~\ref{algo:GPUlabelfseeddedge}}
    \State Label extra seed and frontier edges \Comment{Algorithm~\ref{algo:labelExtraFrontierEdge}}
    \State Change attributes \Comment{Algorithm~\ref{algo:GPUChangeAtributes}}
    \State Search frontier edges for each seed edge \Comment{Algorithm~\ref{algo:Searchfrontieredges}}
    \State Overwritte seeds \Comment{Algorithm~\ref{algo:Overwritteseeds}}
    \State Scan and compact seed edges array
  \end{algorithmic}
\end{algorithm}




\subsection{GPU longest edge labeling}\label{subsec:label longest kernel}


GPolylla starts by calculating the longest-edge of each triangle, this is made to define the border of each terminal-edge region $R_i$ and the seeds that will be use to access to polygon of $\tau'$. 

The kernel  to calculating of the longest-edge is shown in Algorithm~\ref{algo:gpu_labelmaxedge}, it is the direct equivalent to the secuential Algorithm showed in Section~\ref{sec:CPUlongesedge}.

For each triangle in $\tau$, the algorithm assign one half-edge $he_i$, that is part of the interior of the triangle, to each thread of the GPU. Then, the kernel calculates length of the half-edge $he_i$, $next(he_i)$ and $prev(he_i)$ and mark which is the longest in the \texttt{longest\_edge\_bitvector}.


\begin{algorithm}
\footnotesize

    \caption{GPU Label phase longest edge labeling}\label{algo:gpu_labelmaxedge}
    
    \begin{algorithmic}[0]
    \Require Triangulation $\tau = (V,E)$
    \Ensure Labeled triangulation $\tau_L = (V,E)$
    
    \BKernel{\textsc{GPU Label longest edge labeling}($\tau$, \texttt{longest\_edge\_bitvector})}
    \ForAll{halfedge $he_i$ in $\tau$ \textbf{in parallel} }\label{algolabel:gpu_labelmaxedge}
        \State $d_1, d_2, d_3 \leftarrow$ length size of $he_i$, $next(he_i)$, $prev(he_i)$
        \State $he_{max} \leftarrow$ $max(d_1, d_2, d_3)$ 
        \State \texttt{longest\_edge\_bitvector}[$he_{max}$] = True
    \EndFor
    \EKernel
    \end{algorithmic}
\end{algorithm}

\subsection{GPU frontier-edges labeling}\label{subsec: GPU label longest kernel}

In this kernel we calculate the border of a terminal-edge region $R_i$ using the \texttt{longest\_edge\_bitvector} created in the previous kernel. This is done by computing if $he_i \in \tau$ is a frontier-edge and it is the direct equivalent to the secuential Algorithm showed in Section~\ref{sec:CPUlabelfrontieredge}. The kernel is showed in~\ref{algo:GPUlabelfrontieredge}.

The algorithm assign a thread to each half-edge $he_i \in \tau$. The kernel uses the \texttt{longest\_edge\_bitvector} to check if $he_i$ is not the longest-edge edge of its incident triangle and to triangle that contains \textsc{twin($he_i$)}. If it is the case, them $he_i$ is a frontier-edge and is mark as true in {\tt frontier-edge bitvector}[$he_i$]. If $he_i$ is a border-edge then it is marked as true.

\begin{algorithm}
\footnotesize

    \caption{Secuential Label phase: Label frontier edges}\label{algo:GPUlabelfrontieredge}
    \begin{algorithmic}[0]
    \Require Triangulation $\tau = (V,E)$
    \Ensure Labeled triangulation $\tau_L = (V,E)$
    \BKernel{\texttt{LabelFrontierEdges}($\tau = (V,E))$}
    \ForAll{halfedge $he_i \in \tau$ \textbf{in parallel}} 
        \State \texttt{is\_not\_longest\_edge?} $\leftarrow$ $he_i$ and twin($he_i$) are not longest\_edges
        \State \texttt{is\_border\_edge?} $\leftarrow$ $he_i$ or twin($he_i$) is a boundary edge
        \If{\texttt{is\_longest\_edge?} or is\_border\_edge?}
            \State Label $he_i$ as frontier-edge
        \EndIf    
    \EndFor
    \EKernel
    \end{algorithmic}
\end{algorithm}



\subsection{GPU seed-edges labeling} \label{subsec:GPUseededges}

In this phase, the algorithm labels those half-edges that are terminal-edges using the \texttt{longest\_edge\_bitvector} in the \texttt{seed\_edge\_bitvector}. The algorithm selects these half-edges as seeds half-edges because there is only one terminal-edge within a terminal-edge region $R_i$, and this edge can be used to generate the polygon after changing the adjacencies of the frontier-edges.

The kernel is shown in Algorithm~\ref{algo:GPUlabelfseeddedge}. For each half-edge $he_i$ in $\tau$, the kernel checks if both $he_i$ and \textsc{twin($he_i$)} are marked as the longest edge in \texttt{longest\_edge} and if neither is a border-edge. This indicates that both are terminal-edges. Conversely, if $he_i$ or \textsc{twin($he_i$)} is marked as the longest edge in \texttt{longest\_edge} and one of them is a border-edge, this indicates they are border terminal-edges. If one condition is true, then the half-edge between $he_i$ and \textsc{twin($he_i$)} is labeled as a seed edge in the resulting bit-vector {\tt seed-bitvector}. 

It's worth noting that the bit-vector {\tt seed-bitvector} is a sparse array that contains zeros and ones, which makes it sub-optimal for the traversal phase. Since we cannot determine the exact number of seed edges in advance, we need to assign one seed edge to each thread during this phase.

\begin{algorithm}
  \footnotesize
  
      \caption{Label seed edge}\label{algo:GPUlabelfseeddedge}
      \begin{algorithmic}[0]
      \Require Triangulation $\tau = (V,E)$
      \Ensure Labeled triangulation $\tau_L = (V,E)$
      \BKernel{\texttt{LabelSeedEdges}($\tau = (V,E))$}
      \ForAll{halfedge $he_i \in \tau$ \textbf{in parallel}} 
          \State \texttt{is\_terminal\_edge?} $\leftarrow$ $he_i$ and twin($he_i$) are max edges and not border 
          \State \texttt{is\_terminal\_border\_edge?}  $\leftarrow$ $he_i$ or twin($he_i$) is max edge and border
          \If { \texttt{is\_terminal\_edge?} or \texttt{is\_terminal\_border\_edge?} }
              \State Label $he_i$ or twin($he_i$) as seed-edge
          \EndIf    
      \EndFor
      \EKernel
      \end{algorithmic}
  \end{algorithm}


\subsection{Label extra seed and frontier edges}\label{subsec:GPUlabelExtraFrontierEdge}

In order to do the repair phase in parallel, we have to do this extra step, in this phase the algorithm convert internal-edges adjacent to a barrier tip to frontier-edges, thus the algorithm avoid generates non-simple polygons in the first place. 

The kernel called in this step is shown in Algorithm~\ref{algo:labelExtraFrontierEdge}, notice that this kernel is the equivalent of the first part of the repair phase showed in Algorithm~\ref{algo:CPURepairphase}. 

For each vertex in $v_i \in \tau = (V,E)$, the kernel count the number of frontier-edges adjacent to $v_i$, if there is only one frontier-edge adjacent to $v_i$, then $v_i$ is a barrier tip, thus the kernel count the number of edges adjacent to $v_i$, and select one of them, the middle one counting from a frontier-edge, and convert the internal-edge to a frontier-edge, labeling their two half-edges as frontier-edges and as seed-edges, thus the algorithm can use them to generate to new polygons.

The part of store two new seed half-edges must be done, as is show in Figure~\ref{figs:splitmid}, a non-simple polygon only have one seed to generate that polygon, but it could generate several new polygons after repair it. All those new polygons needs a seed to be generated too, thus the algorithm needs to store multiple new seeds, the problem with this addition, is one polygon could have more of one seed. But it will be fixed in the kernel of the subsection~\ref{subsec:GPUoverwritteseeds}.

\begin{algorithm}
  \footnotesize
  \caption{Label Extra Frontier Edge}\label{algo:labelExtraFrontierEdge}
  \begin{algorithmic}[0]
    \Require Labeled $\tau_L = (V,E)$
    \Ensure Updated $frontier\_edges$ and $seed\_edges$
    \BKernel{\texttt{label\_extra\_frontier\_edge\_d}($\tau = (V,E)$)}
    \ForAll{vertices $v_i \in \tau$ \textbf{in parallel}} 

      \State $he$  $\leftarrow$ \textsc{edgeOfVertex($v_i$)}
      \State numFrontierEdges $\leftarrow$ 0
      \Repeat \Comment{Count frontier-edges adjacent to $v_i$}
        \If{$he$ is a frontier-edge} 
          \State numFrontierEdges $\leftarrow$ numFrontierEdges + 1 
        \EndIf
        \State $he \leftarrow$ \textsc{CWvertexEdge($he$)} 
      \Until{$he$ is not \textsc{edgeOfVertex($v_i$)}}
      \If{numFrontierEdges is equal to 1} \Comment{If $v_i$ is barrier tip}
          \State $he \leftarrow$ \textsc{edgeOfVertex($v_i$)} 
          \While{$he$ is not a frontier-edge} \Comment{Find middle edge}
            \State $he \leftarrow$ \textsc{CWvertexEdge($he$)}
          \EndWhile
          \For{0 to $(\textsc{degree}(v_i)-1)/2$}  \Comment{set $he$ as middle edge}
            \State $he \leftarrow$ \textsc{CWvertexEdge($he$)}
          \EndFor 
          \State Label $he$ and \textsc{twin}($he$) as frontier-edge
          \State Label $he$ and \textsc{twin}($he$) as seed-edges
        \EndIf
        \EndFor
    \EKernel
  \end{algorithmic}
\end{algorithm}

\subsection{Change attributes} \label{subsec:GPUchangeattributes}

After labeling the frontier-edges and seed-edges, the algorithm can start the process of changing the adjacencies of the attributes of each half-edges showed in Algorithm~\ref{algo:CPUtravelphase}. The kernel that do this is showed in Algorithm~\ref{algo:GPUChangeAtributes}. 

For each half-edge $he_i \in \tau$ in parallel, the kernel search the next and previous frontier-edge, using the same idea showed in Figure~\ref{fig:traversalphase}, but in the case of the search of the previous frontier-edge, the algorithm uses the \textsc{CCWvertexEdge} query instead of the \textsc{CWvertexEdge($he$)}.


\begin{algorithm}
  \footnotesize
  
      \caption{GPU Change atributes kernel}\label{algo:GPUChangeAtributes}
      \begin{algorithmic}[0]
      \Require Labeled triangulation $\tau_L = (V,E)$
      \Ensure Polylla mesh 
      \BKernel{\texttt{Traversal phase}($\tau_L = (V,E)$)}
          \ForAll{halfedge $he_i$ in $\tau_L$ \textbf{in parallel} }

                  \State next $\leftarrow$ $he_i$

                  \While{next is not a frontier-edge} \Comment{Search next frontier-edge in CW}
                    \State next $\leftarrow$ \textsc{CWvertexEdge($he$)}
                  \EndWhile

                  \State \texttt{set\_next(next($he_i$))} $\leftarrow$ next
                  \State prev $\leftarrow$ $he_i$
                  \While{prev is not a frontier-edge} \Comment{Search prev frontier-edge in CCW}
                    \State prev $\leftarrow$ \textsc{CCWvertexEdge($he$)}
                  \EndWhile
                  \State \texttt{set\_prev(prev($he_i$))} $\leftarrow$ prev

          \EndFor
      \EKernel

    \end{algorithmic}
\end{algorithm}

\subsubsection{Search frontier edges for each seed edge} \label{subsec:GPUsearchfrontieredges}

At this point, the algorithm already have a Polylla mesh in the half-edge data structure, but there are two problems, some seeds are internal-edges of the triangulation and not half-edges of the Polylla mesh, and there are some polygons with more of one seeds to generate it. In this kernel we are going to solve the first problem.

The reason of why there are seed that are internal edges is because when the algorithm does the process of label the seed-edges we choose one of the two half-edges of the terminal-edges as a seed half-edge. In the kernel showed in Algorithm~\ref{algo:Searchfrontieredges} for each seed half-edge the algorithm search a new frontier-edge using the operation \textsc{CWvertexEdge} and after find one, the algorithm remove the label of the seed half-edge and label the found frontier-edge as a new seed edge. 

After this kernel the algorithm only need to remove extra seed half-edges to complete the Polylla mesh.

\begin{algorithm}
  \footnotesize
  
      \caption{GPU Search frontier edges for each seed edge}\label{algo:Searchfrontieredges}
      \begin{algorithmic}[0]
      \Require Labeled triangulation $\tau_L = (V,E)$
      \Ensure Polylla mesh 
      \BKernel{\texttt{Traversal phase}($\tau_L = (V,E)$)}
          \ForAll{halfedge $he_i$ in $\tau_L$ \textbf{in parallel} }
            \If{$he_i$ is a seed edge} 
                  \State next $\leftarrow$ $he_i$

                  \While{next is not a frontier-edge} \Comment{Search next frontier-edge in CW}
                    \State next $\leftarrow$ \textsc{CWvertexEdge($he_i$)}
                  \EndWhile
                  \If{next is not $he_i$}
                      \State \texttt{seed\_edge\_bitvector}[$he_i$] = \texttt{False} \Comment{Set the frontier-edge a a seed edge}
                  \EndIf    
                  \State \texttt{seed\_edge\_bitvector}[$he_i$] = \texttt{True} \Comment{Remove the original seed edge}
            \EndIf
          \EndFor
      \EKernel

  \end{algorithmic}
\end{algorithm}

\subsection{Overwritte seeds} \label{subsec:GPUoverwritteseeds}

Note that in the kernel presented in subsection~\ref{subsec:GPUlabelExtraFrontierEdge}, the algorithm labels two adjacent half-edges to an interior-edge as frontier-edges to split a non-simple polygon. To ensure that the new polygons have a seed, the algorithm also labels both half-edges as seed edges. However, this poses a problem because the algorithm might generate the same polygon twice. To prevent this, the algorithm overwrites the seed half-edges using the kernel presented in Algorithm~\ref{algo:Overwritteseeds}.

For each seed half-edge, the kernel traverses inside the polygon that generated the seed. During this traversal, it searches for the frontier half-edge with the smallest index and labels this half-edge as a seed half-edge. To avoid race conditions, the algorithm first checks if the minimum index $min$ is not equal to the index of the original seed $i$. If this is true, the algorithm sets \texttt{seed\_edge\_bitvector}[$i$] to False. Subsequently, it sets \texttt{seed\_edge\_bitvector}[$min$] of the frontier-edge with the minimum index to True.


It is important to note that this kernel is also implemented in the sequential algorithm, specifically in the repair phase as outlined in~\ref{sec:CPURepairPhase}. The algorithm~\ref{algo:CPURepairphase} is divided into two kernels: the first part is executed in Kernel~\ref{subsec:GPUlabelExtraFrontierEdge}, and the second part is executed in the kernel described in this subsection.




\begin{algorithm}
  \footnotesize
  
      \caption{GPU Overwritte seeds}\label{algo:Overwritteseeds}
      \begin{algorithmic}[0]
      \Require Labeled triangulation $\tau_L = (V,E)$
      \Ensure Polylla mesh 
      \BKernel{\texttt{Traversal phase}($\tau_L = (V,E)$)}
          \ForAll{halfedge $he_i$ in $\tau_L$ \textbf{in parallel} }

          \State ${init}$ $\leftarrow$ $he_i$
          \State ${min}$ $\leftarrow$ ${init}$
          \State ${curr}$ $\leftarrow$ next(init)
              \While {init is not curr} 
                \State min $\leftarrow$ minimum(min, curr)    
                \State curr $\leftarrow$ next(curr)
              \EndWhile
              \If{min is not $he_i$}
                \State \texttt{seed\_edge\_bitvector}[$he_i$] = \texttt{False}
              \EndIf    
              \State \texttt{seed\_edge\_bitvector}[${min}$] = \texttt{True}
          \EndFor
      \EKernel

  \end{algorithmic}
\end{algorithm}

\subsection{Scan and compact seed edges array} \label{subsec:GPUscanandcompactseededgesarray}

Until this kernel we already have a polylla mesh generated, but the output seed edges are in a bitvector and to be converted to integer array. This is done with the classical scan and compact technique, but accelerate with tensor cores

In this kernel the algorithm used a technique based on prefix sum to compact the seed-edges array, which was also used in~\cite{carrasco2023evaluation}. This technique consists of computing the prefix sum of all elements of the seed-list bit-vector, this sum says the location in the compacted output array. 

This step is not equivalent to any step in the secuential algorithm as in the secuential algorithm the seed edges are stored in a list, but in the GPU algorithm the seed edges are stored in a bit-vector, thus the algorithm needs to convert the bit-vector to a list.

\section{Experiments}\label{sec:experiments}

This section describes the experimentation conducted in this study, including the dataset utilized, tests carried out (experiments and benchmark environment), and the results obtained.

\subsection{Dataset}

\begin{figure}[h]
  
  \begin{subfigure}[b]{0.44\textwidth}
  \centering
    \includegraphics[width=0.7\textwidth]{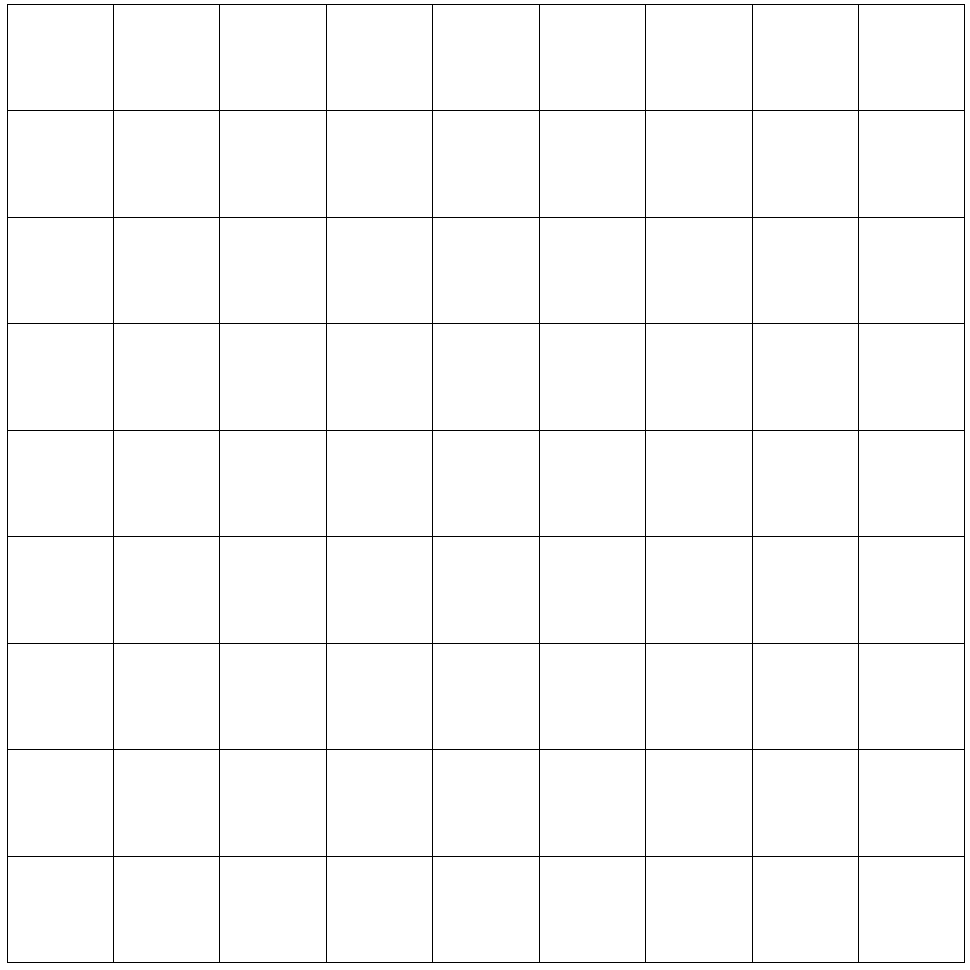}
    \caption{Uniform Distribution}
    \label{fig:subfig_a}
  \end{subfigure}
  \begin{subfigure}[b]{0.44\textwidth}
  \centering
    \includegraphics[width=0.7\textwidth]{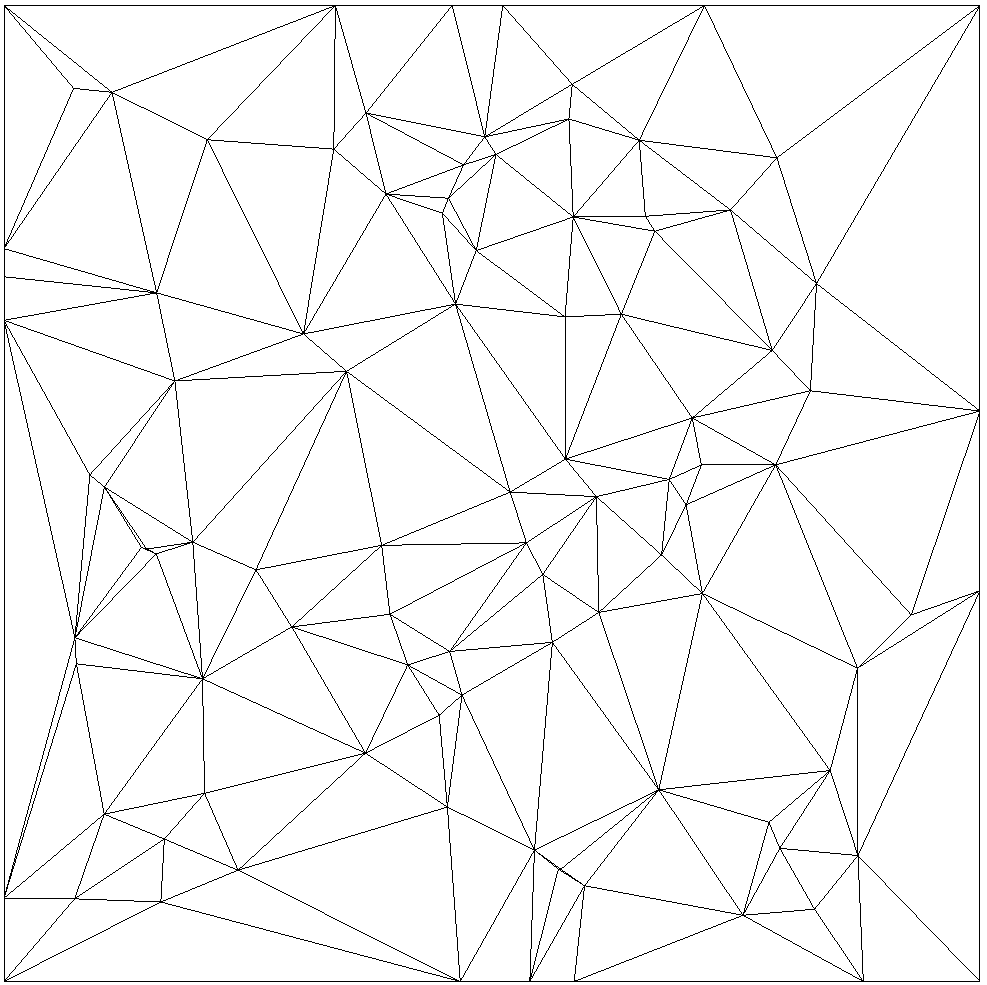}
    \caption{Delaunay distribution}
    \label{fig:subfig_b}
  \end{subfigure}
  \caption{The illustration shows both input tests of our experiment, the left one is a uniform grid with the 100 closest equidistant square roots perfect, and on the right is an example of 100 points on a Delaunay distribution.}
  \label{fig:two_subfigures}
\end{figure}

The Figure~\ref{fig:two_subfigures} shows two different point distributions that we used to test out our algorithm. The first one, in Figure~\ref{fig:subfig_a}, is a totally uniform grid, this kind of grid is generated using the Algorithm~\ref{algo:gen_points}. The second one is shown in Figure~\ref{fig:subfig_b}, it is a Delaunay mesh generate using random uniform points, this kind of mesh is generated by randomly placing points on a square, without overlapping points, and using a tolerance parameter $\delta$ to move the points that are near to the border of the square to the border, afterward using the software Triangle~\cite{triangle2d} to generate a Delaunay mesh. For the rest of the experiments we will call to the first kind of mesh a \emph{Grid meshes} and to the second a \emph{Random meshes}.

\begin{algorithm}
\footnotesize

\caption{Generate points and triangles for a 2D mesh}
\label{algo:gen_points}
\begin{algorithmic}[0]
\Require $n > 1$
\Ensure List of vertices and triangle indices for a 2D mesh with $n$ points
\Procedure{GenerateMesh}{$n$}
\State Create an empty list of vertices $Vertices$
\State Create an empty list of triangle indices $faces$
\State $sqrt\_n \gets \lfloor \sqrt{n} \rfloor$
\ForAll{$i$ from $0$ to $\text{sqrt\_n}-1$}
\ForAll{$j$ from $0$ to $\text{sqrt\_n}-1$} \label{algolabel:generatevertices}
\State Create a new vertex $ve$ with coordinates $(i,j)$
\State Add $ve$ to the list $Vertices$
\EndFor
\EndFor
\ForAll{$i$ from $0$ to $n-\text{sqrt\_n}-1$} \label{algolabel:generatetriangles}
\If{$i \bmod \text{sqrt\_n} \neq \text{sqrt\_n}-1$}
\State Add $i$ to the list $faces$
\State Add $i+1$ to the list $faces$
\State Add $i+\text{sqrt\_n}+1$ to the list $faces$
\State Add $i$ to the list $faces$
\State Add $i+\text{sqrt\_n}+1$ to the list $faces$
\State Add $i+\text{sqrt\_n}$ to the list $faces$
\EndIf
\EndFor
\State \textbf{return} $Vertices$, $faces$
\EndProcedure
\end{algorithmic}
\end{algorithm}

\subsection{Experimental setup}
Our implementation employed C++ with -O3 optimization for the CPU component and CUDA with NVCC 12 for the GPU component. We conducted all of our experiments on the Patagón supercomputer \cite{patagon}, which is equipped with a single Nvidia DGX A100 GPU node, two  AMD EPYC 9534 CPUs with 64 cores and 256MB L3 cache, 756 GB of RAM DDR5, and 3 Nvidia L40 GPUs, each with 48GB of VRAM GDDR with ECC. However, for the purposes of our experiments, we utilized only a single L40 GPU.

The present work explores and compares the effectiveness of different point distributions through the implementation of two experiments. The first experiment involved utilizing a Delaunay distribution with 32 equidistant intervals ranging from one million to 46 million points. The second experiment employed a uniformly distributed grid with the 32 nearest roots to the equidistant intervals between one million and 100 million. These experiments were limited by the current memory constraints of graphics cards, as the author attempted to reach the maximum number of points that their current hardware could support. However, it is important to note that there are no such limitations at the programming level, and it is possible to scale the number of points as new hardware with higher capacity becomes available.

\subsection{Results}

\begin{table*}
\centering
\resizebox{1\columnwidth}{!}{%
\begin{tabular}{lcccccccccccccccccc}
  \toprule
  {} & \multicolumn{6}{c}{CPU} & \multicolumn{12}{c}{GPU} \\
  \cmidrule(lr){2-7} \cmidrule(lr){8-19}
  \#V &      LM &      LF &      LS &    Trav &    Rep &   Total &   CtD &  LLK &  LFK &  LSK &  LEK &  CaK & SFK &   BtH &  OSK & Scan &   TwC & Total \\
  \midrule
  \textbf{1M } &   917.0 &   256.0 &   333.8 &   361.1 &   43.8 &  1911.8 &   7.0 &  0.4 &  0.2 &  0.2 &  1.2 &  0.4 & 0.2 &   8.8 &  0.3 &  0.3 &  19.0 &   3.2 \\
  \textbf{12M} & 11246.7 &  3037.6 &  3943.3 &  4274.7 &  538.4 & 23040.8 &  80.7 &  5.4 &  2.8 &  2.7 & 25.8 &  4.6 & 2.2 & 100.0 &  3.1 &  1.8 & 229.1 &  48.4 \\
  \textbf{23M} & 21727.2 &  5817.2 &  7555.6 &  8263.2 & 1046.8 & 44410.1 & 153.7 & 10.3 &  5.3 &  5.2 & 50.3 &  8.8 & 4.1 & 195.7 &  5.8 &  3.3 & 442.5 &  93.2 \\
  \textbf{33M} & 32300.9 &  8590.3 & 11173.1 & 12165.1 & 1549.6 & 65779.0 & 227.3 & 15.2 &  7.9 &  7.6 & 74.9 & 13.0 & 6.0 & 281.0 &  8.6 &  4.7 & 646.1 & 137.9 \\
  \textbf{44M} & 42827.9 & 11451.2 & 14751.8 & 16122.6 & 2046.5 & 87200.0 & 299.9 & 20.0 & 10.4 & 10.0 & 99.2 & 17.2 & 8.0 & 419.4 & 11.4 &  6.2 & 901.7 & 182.4 \\
  \midrule
  \textbf{1M  } &   675.5 &   178.8 &   274.0 &   303.8 & 0.0 &   1432.0 &   7.1 &  0.4 &  0.2 &  0.2 &  0.4 &  0.3 &  0.2 &    9.0 &  0.2 &  0.3 &   18.3 &   2.2 \\
\textbf{27M } & 18215.8 &  4855.4 &  7341.4 &  8241.9 & 0.0 &  38654.5 & 179.7 &  8.8 &  6.4 &  6.5 &  8.3 & 10.1 &  4.3 &  237.9 &  4.8 &  4.2 &  470.9 &  53.3 \\
\textbf{49M } & 33451.3 &  8912.5 & 13485.0 & 15180.5 & 0.0 &  71029.3 & 332.1 & 15.9 & 11.8 & 11.9 & 15.1 & 18.6 &  7.8 &  436.8 &  8.9 &  7.5 &  866.5 &  97.6 \\
\textbf{74M } & 51456.9 & 13746.1 & 20822.9 & 23551.0 & 0.0 & 109577.0 & 522.0 & 24.1 & 18.1 & 18.2 & 22.9 & 28.4 & 11.8 &  876.9 & 13.5 & 11.4 & 1547.2 & 148.3 \\
\textbf{100M} & 68806.4 & 18513.8 & 27816.8 & 31814.4 & 0.0 & 146951.4 & 741.0 & 32.1 & 24.4 & 24.6 & 30.3 & 38.1 & 15.8 & 1027.5 & 18.1 & 15.2 & 1967.0 & 198.4 \\
  \bottomrule
  \end{tabular}  

}
\caption{Time measurements for the CPU and GPU versions of Polylla. The times are in miliseconds. The upper table is the Random meshes and the lower table is the Grid meshes. The table presents the number of vertices (\#V) for each mesh, along with the timings for different stages of the algorithm. For the CPU version: "Label the longest-edge" (LM), "Label frontier edges" (LF), "Label seed edges" (LS), "Traversal phase" (Trav), and "Repair phase" (Rep), cumulating in the total time for CPU Polylla (Total). The GPU version encompasses: "Copy to Device" (CtD), "Label the longest-edge kernel" (LLK), "Label frontier edges kernel" (LFK), "Label seed edges kernel" (LSK), "Label extra seed and frontier edges kernel" (LEK), "Change attributes" (CaK), "Search frontier edges for each seed edge" (SFK), "Overwrite seeds" (OSK), "Scan and compact seed edges array" (Scan), "Copy back to Host" (BtH), culminating in the total time for GPU Polylla excluding copy times (Total).}
\label{tab:time_CPU_and_GPU}
\end{table*}

\begin{figure}[ht]
  \centering
  \begin{subfigure}[b]{0.47\textwidth}
    \includegraphics[width=\textwidth]{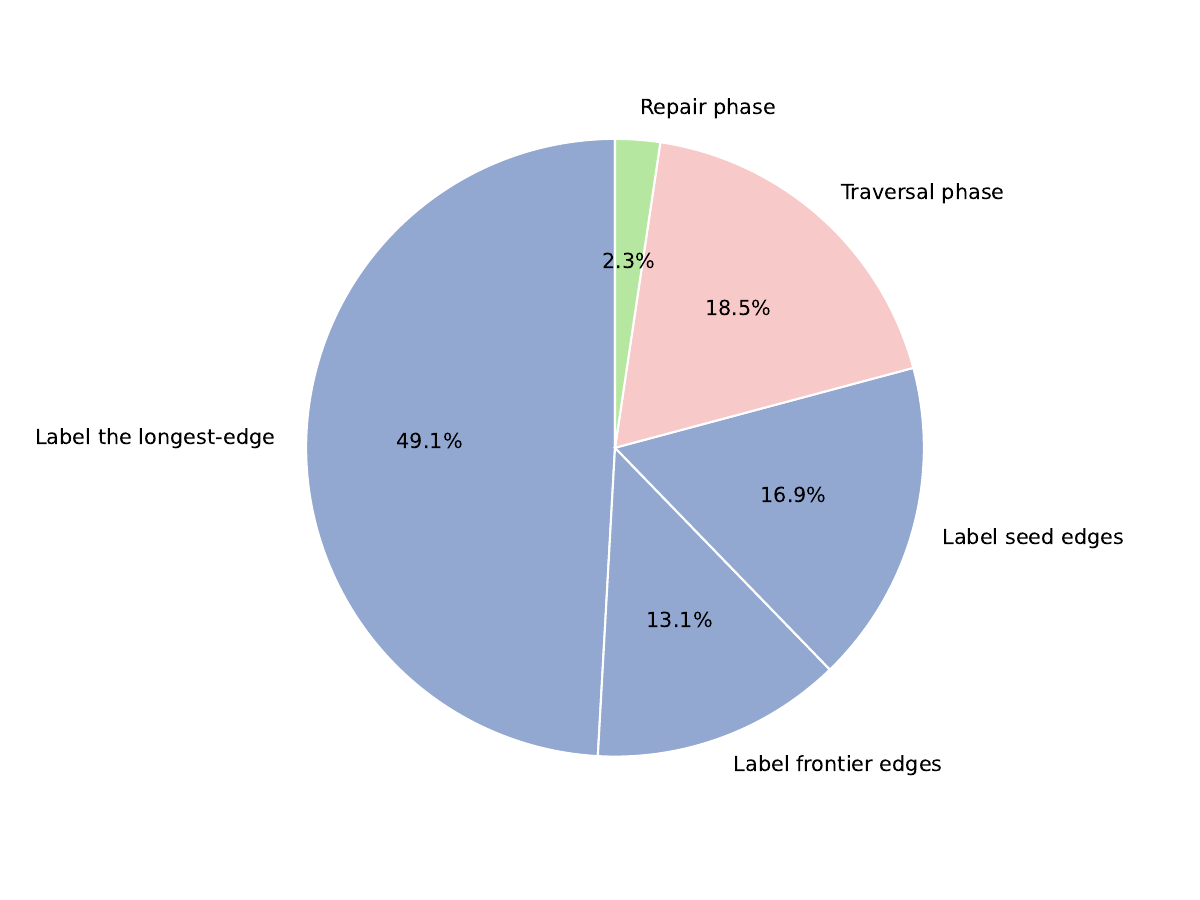}
    \caption{CPU random meshes}
    \label{fig:CPU_pie_chart_random}
  \end{subfigure}
  \hfill 
  \begin{subfigure}[b]{0.47\textwidth}
    \includegraphics[width=\textwidth]{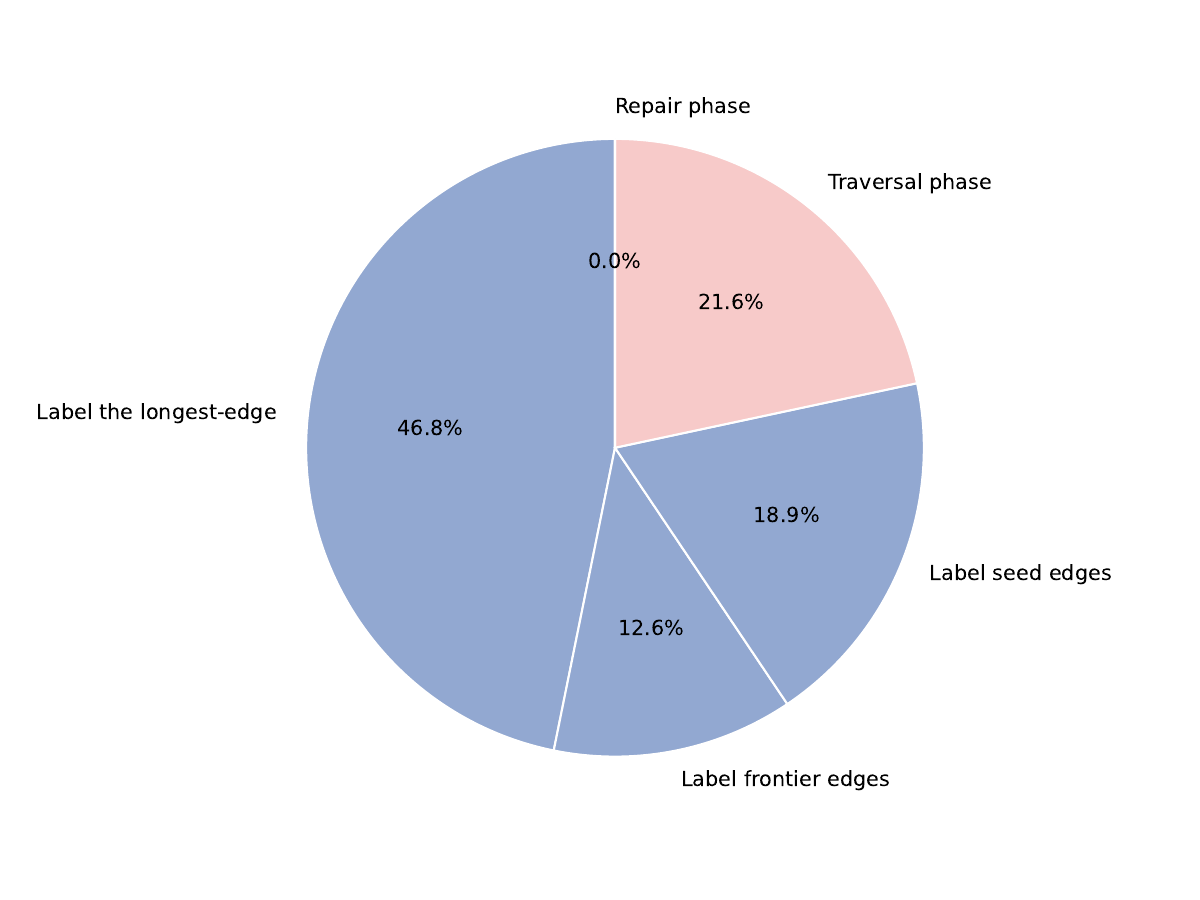}
    \caption{CPU grid meshes}
    \label{fig:CPU_pie_chart_grid}
  \end{subfigure}
  
  \vspace{1em} 
  
  \begin{subfigure}[b]{0.47\textwidth}
    \includegraphics[keepaspectratio, width=\textwidth]{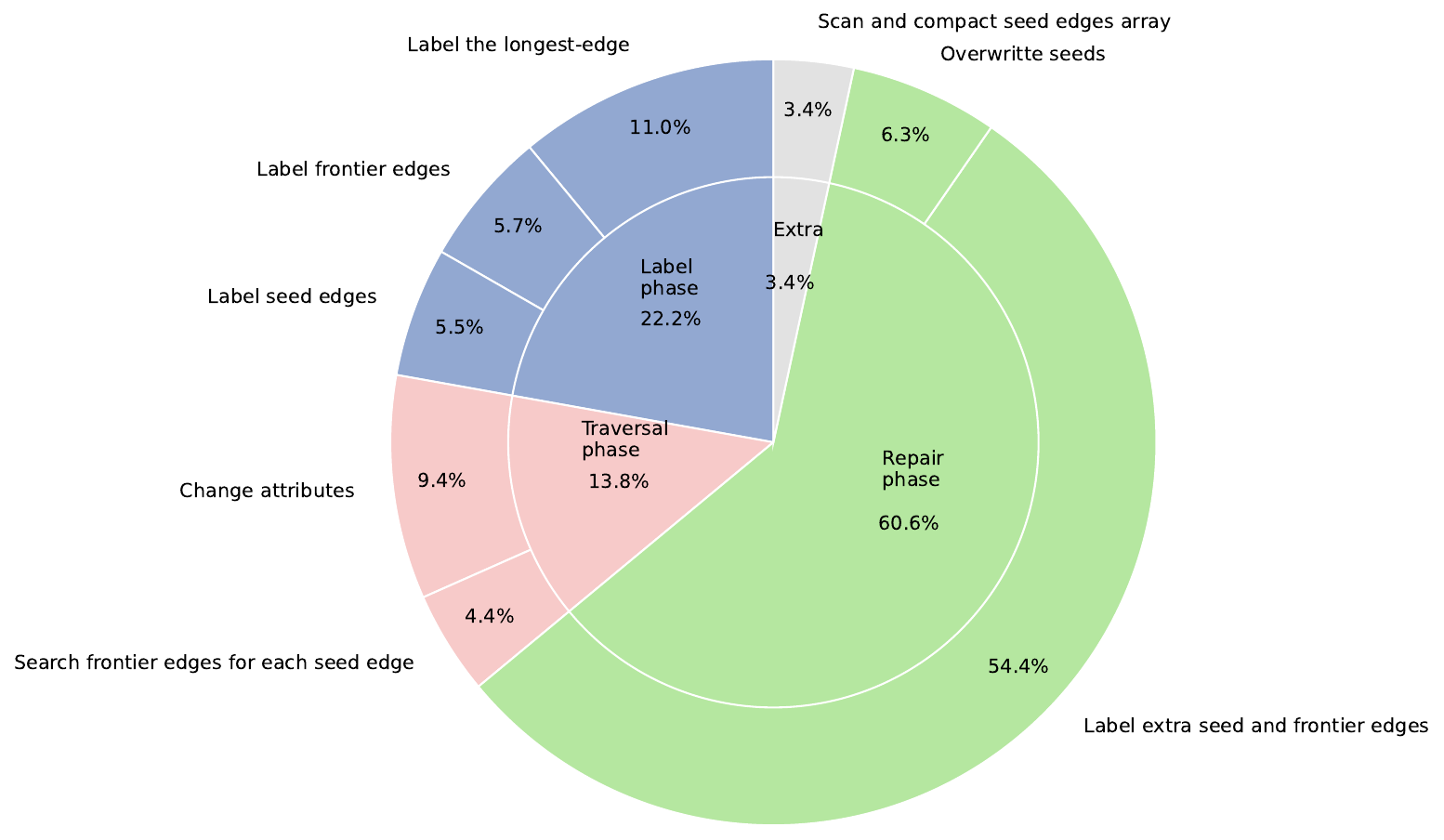}
    \caption{GPU random meshes}
    \label{fig:GPU_pie_chart_random}
  \end{subfigure}
  \hfill 
  \begin{subfigure}[b]{0.47\textwidth}
    \includegraphics[keepaspectratio,width=\textwidth]{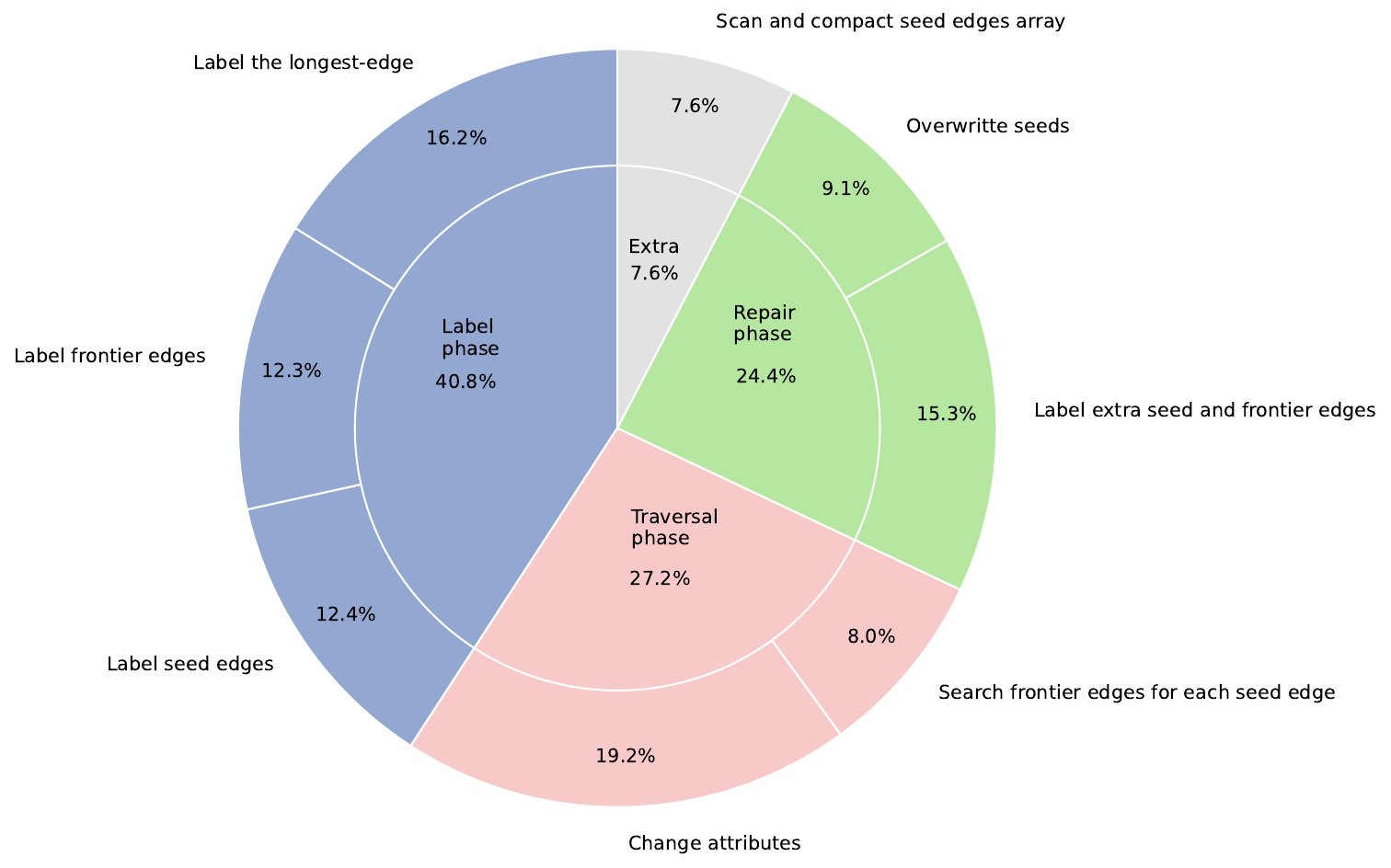}
    \caption{GPU grid meshes}
    \label{fig:GPU_pie_chart_grid}
  \end{subfigure}

  \caption{Comparison of CPU and GPU Performance of each phase of the algorithm. In the case of random meshes we present the result for 44 millon points case, for the grid meshes we present we 100 millons case.}
  \label{fig:pieplot}
\end{figure}

\begin{figure}[ht]
  \centering
  \begin{subfigure}[b]{0.47\textwidth}
    \includegraphics[width=\textwidth]{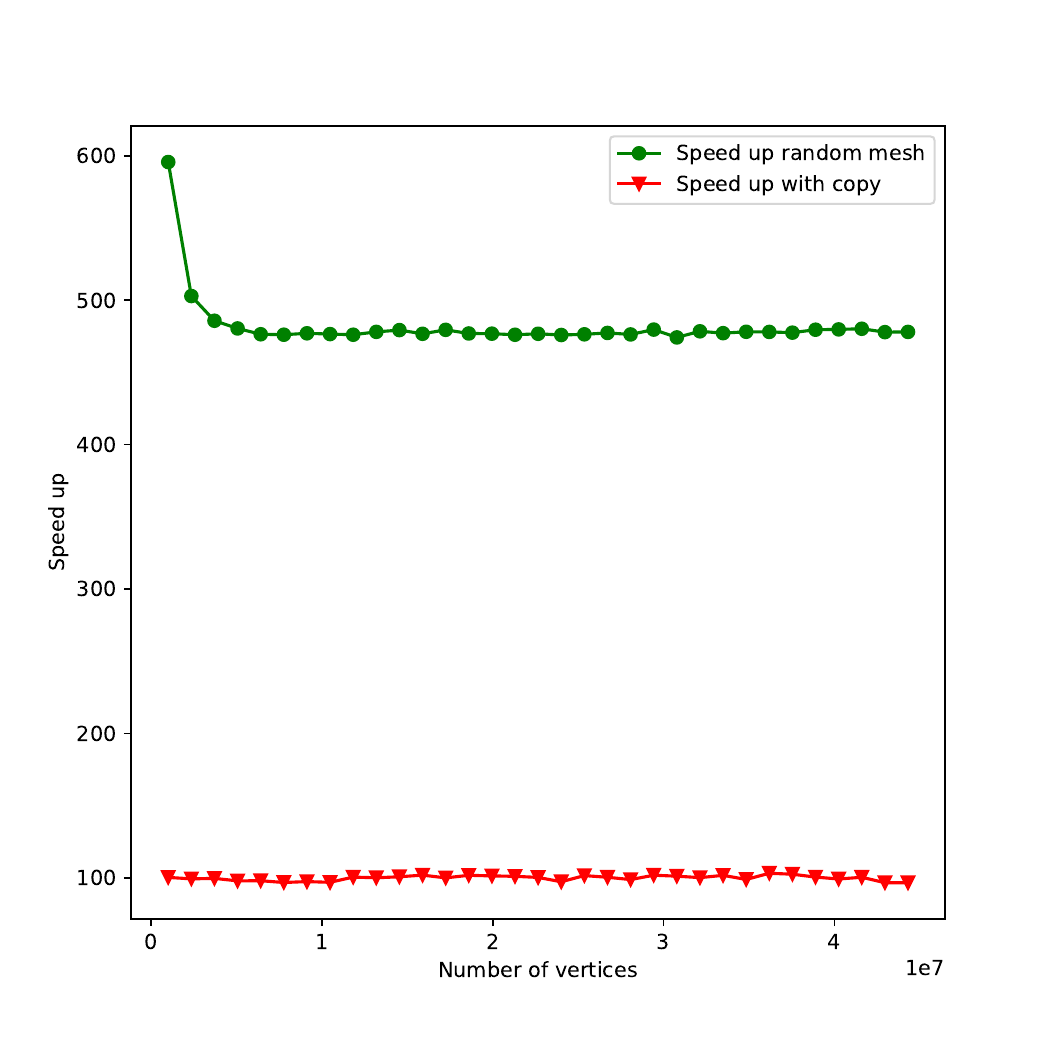}
    \caption{Random meshes}
    \label{fig:speedup_random}
  \end{subfigure}
  \hfill
  \begin{subfigure}[b]{0.47\textwidth}
    \includegraphics[width=\textwidth]{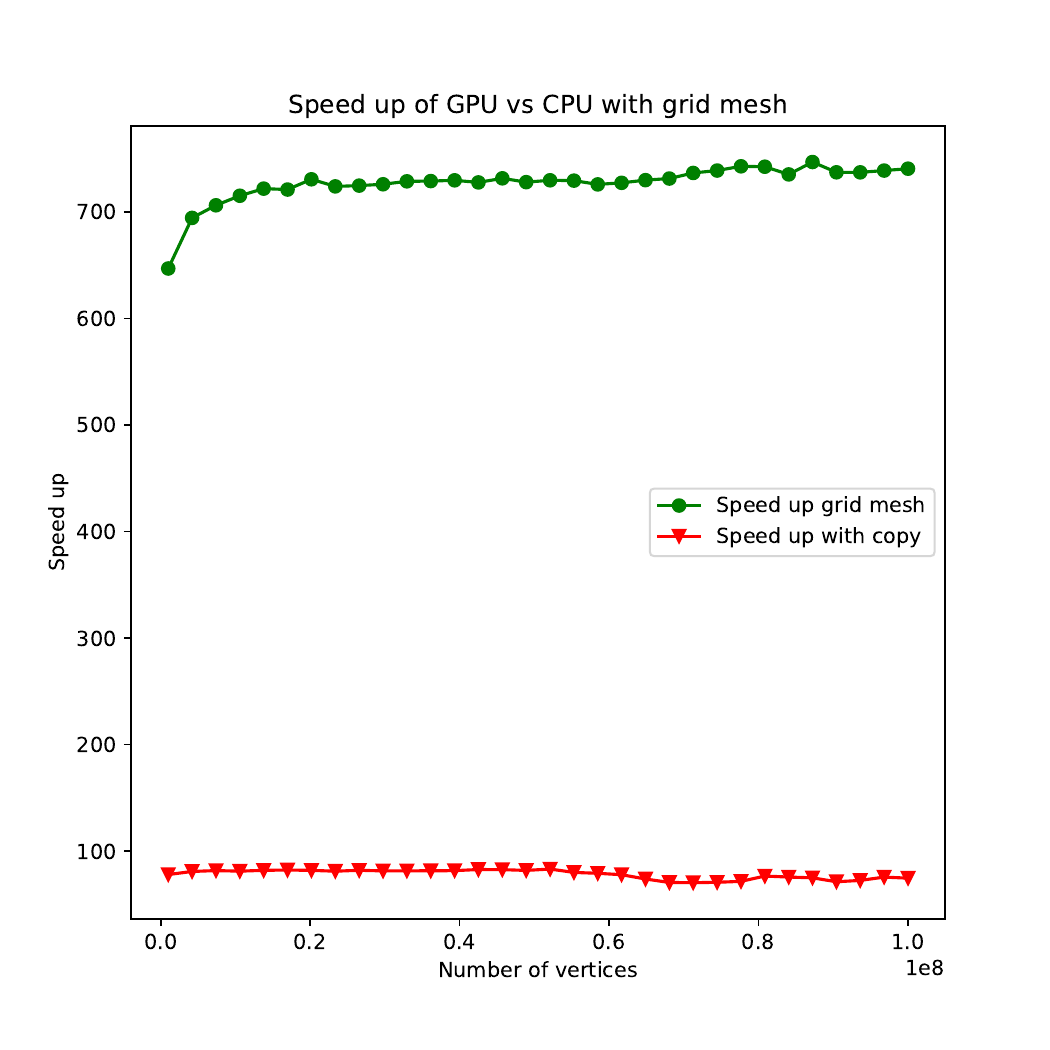}
    \caption{Grid meshes}
    \label{fig:speedup_grid}
  \end{subfigure}
  \caption{The illustration shows both input tests of our experiment, the left one is a uniform grid with 32 closest equidistant perfect square roots, and on the right is an example of 32 points on a Delaunay distribution.}
  \label{fig:speedup}
\end{figure}

\begin{figure}[ht]
  \centering
  \begin{subfigure}[b]{0.47\textwidth}
    \includegraphics[width=\textwidth]{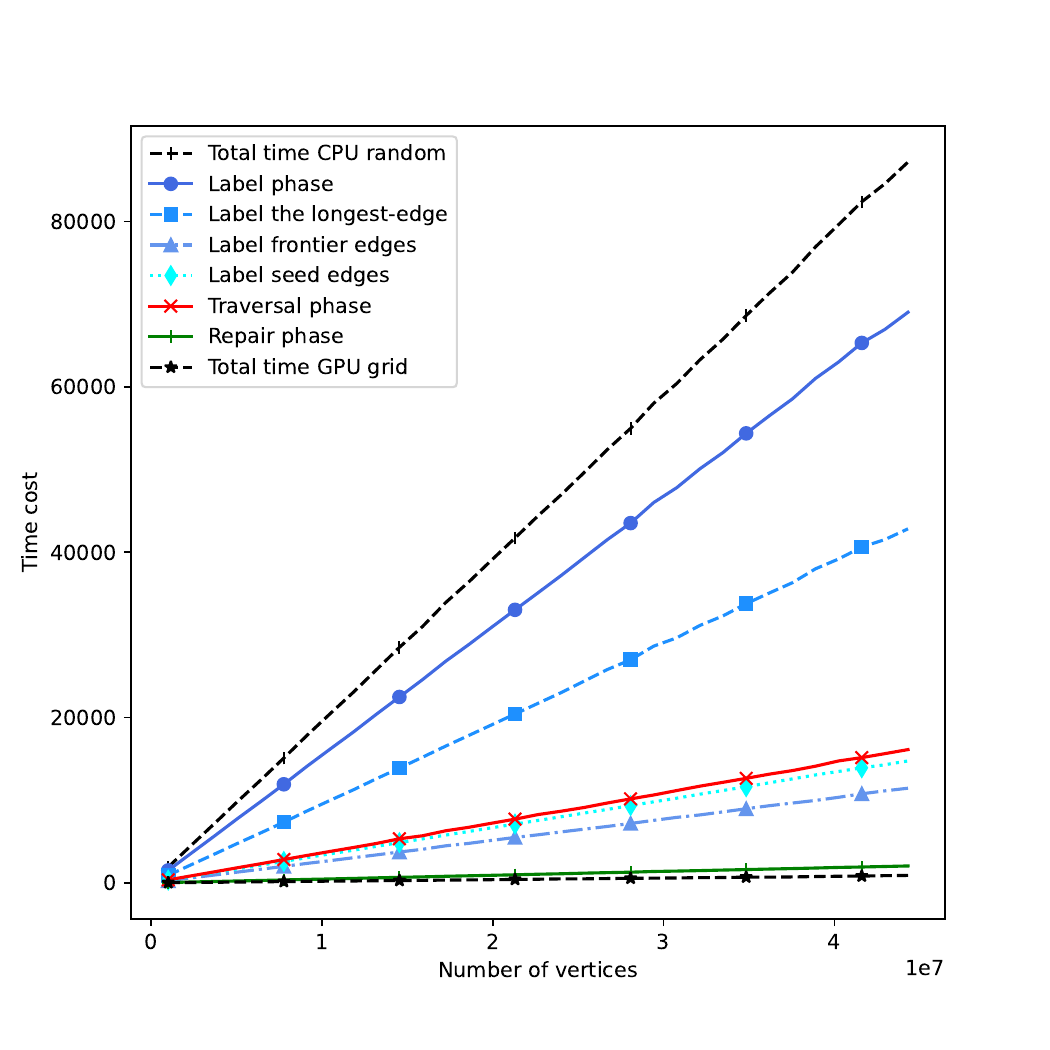}
    \caption{Random meshes}
    \label{fig:CPU_line_plot_random}
  \end{subfigure}
  \hfill
  \begin{subfigure}[b]{0.47\textwidth}
    \includegraphics[width=\textwidth]{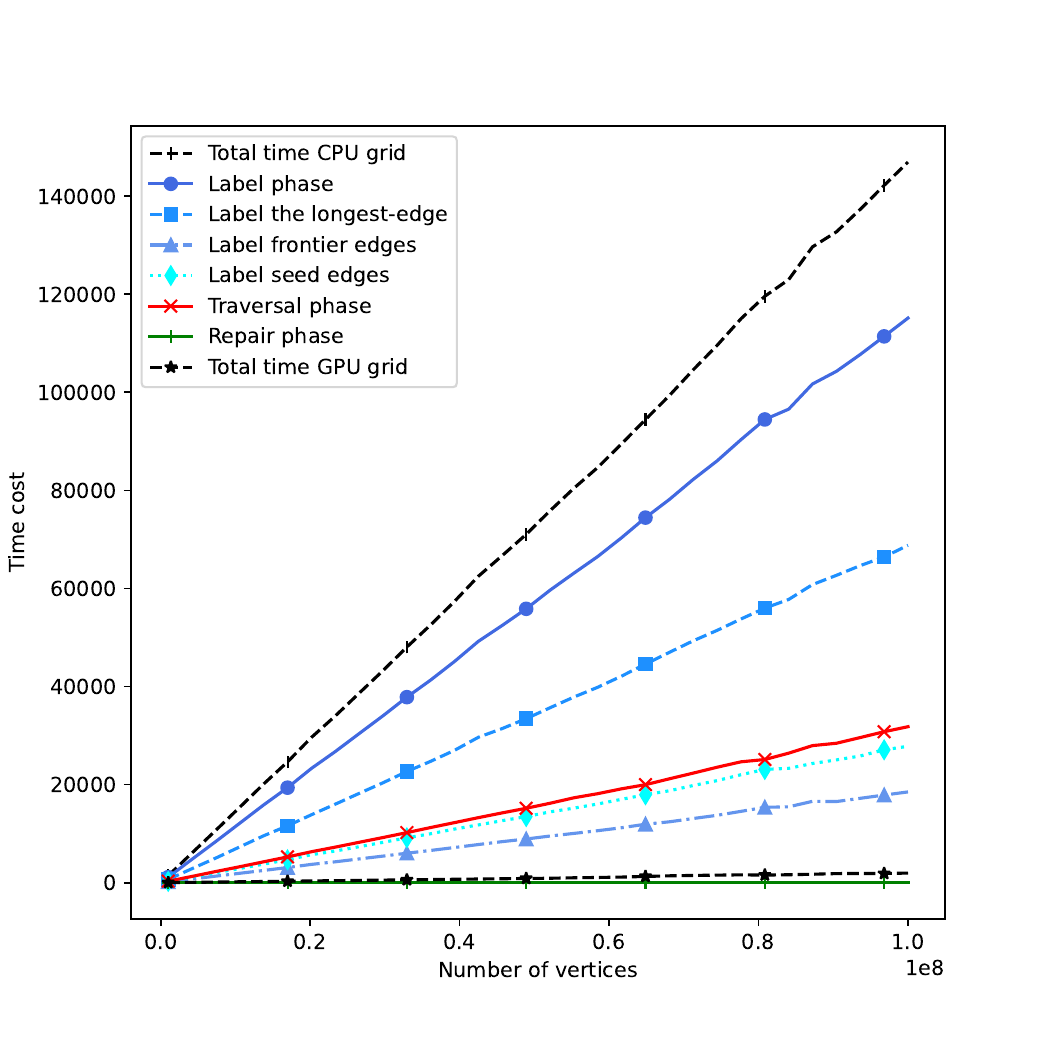}
    \caption{Grid meshes}
    \label{fig:CPU_line_plot_grid}
  \end{subfigure}
  \caption{The illustration shows both input tests of our experiment in CPU and GPU including the repair phase, the left one is a uniform grid with 32 closest equidistant perfect square roots, and on the right is an example of 32 points on a Delaunay distribution.}
  \label{fig:CPU_perfomance}
\end{figure}

Table~\ref{tab:time_CPU_and_GPU} shows the duration of each phase in our CPU and new GPU implementation, along with the time taken for copying to the device for 5 of the 32 experiments of the Random meshes and Grid meshes. The rest of the table can be seen in Appendix~\ref{appendix:tables}. We can see that the maximum speed up the Random meshes is of $\times 595.6$, and if we considerate the copy is of $\times 103.2$ In the case of the Grid meshes the maximum speed up is of $\times 746.8$ and $\times 83.2$ with copy. Those speed up are show in the Figure~\ref{fig:speedup}. Notice that copy the data structure from host to device and copy from device to host is a $79\%$ in average of the total time of the algorithm.

To compare the time cost of each phase in the secuential algorithm and the cost of each kernel in the GPU, we do a pie chart with last rows of the Table~\ref{tab:time_CPU_and_GPU}. In those pie chart we define the kernels of the equivalent phase in the secuential, the only kernel without equivalent is the scan as it is only necessary in the GPU due to the uses of a bitvector to store the output seeds. The charts we can see that kernel related to the repair phase are the most time cost of the algorithm, and even when there are no barrier tips to repair, as it is case of the Grid meshes, there is a cost in the kernel ``Label extra seed and frontier edges" as in this kernel, for each vertex in the triangulation, the algorithm have to check if the vertex is a barrier tip or is not. The ``Overwrite seed kernel" will also have a cost as in this phase we the algorithm set the edge with the minimum index as an output seed edge, when there are barrier tip this aviod to have repited polygon, but when there are not barrier tip, this kernel is no necessary. In the case of the Label phase, the label longest-edge is the most time cost step in both CPU and GPU, but the GPU acceration help to avoid made this process the most costly process in all the algorithm for the Random meshes. 

Finally, to compare to present the magnitude of the low time cost of the GPU Polylla in comparision with the secuential Polylla, the Figure~\ref{fig:CPU_perfomance} show as the GPU Polylla algorithm is even lower in time than the secuential repair phase. In Figure~\ref{fig:CPU_line_plot_grid}, the GPU algorithm lost againts the repair phase only becouse there is no repair phase in those meshes, thus the time cost is $0$.




\clearpage

\section{Conclusions and ongoing work}\label{sec:conclusions}


In this work, we showed a novel way to generate polygonal meshes in GPU, the way that we modify the attributes of each half-edge structure to simulate edge removition and join faces polygonal faces, can be used in the future to accelerate the process that requires mesh simplification in GPU, as can be low polygon mesh generation. Notice that this work is a fully GPU algorithm, the only need of CPU time is to generate the data structure and send it to the GPU.

In conclusion, our algorithm running on a GPU has proven to be highly scalable, allowing for larger meshes to be processed simply by increasing the available graphics memory. We can get a maximum speed up of $\times 746.8$, and if we consider a more realistic case, as a meshes with points in general position (the random meshes) and considering the copy time from device to host and host to device, we can get a speed up of  $\times 83.2$.

As the generations of GPUs progress, the graphic's memory continues to increase, enabling even larger meshes to be processed. Additionally, the performance of the algorithm can be improved by enhancing the speed of transfer between the GPU and CPU, which accounts for 79\% of the algorithm.

Future work for the GPolylla algorithm is using the compact data structure presented in our previous work~\cite{salinas2023generation} to get even bigest meshes, but with one reduction of the speed up.


\section{Acknowledgment}

This research was supported by the Patag\'on supercomputer~\cite{patagon-uach} of Universidad Austral de Chile (FONDEQUIP EQM180042). This work was partially funded by ANID doctoral scholarship 21202379 (first author), ANID doctoral scholarship 21210965 (second author), and ANID FONDECYT grant 1211484 (third author).

\bibliography{sn-bibliography}


\appendix
\section{Tables}\label{appendix:tables}

Here we have the tables used to generate the plots in section~\ref{sec:experiments}. Table~\ref{Fig:table_random_Total} is the table with the CPU and GPU times of the random meshes, and Table~\ref{Fig:table_grid_Total} of the Grid meshes.

\begin{table*}
  
  \resizebox{1\columnwidth}{!}{%
\begin{tabular}{lcccccccccccccccccc}
  \toprule
  {} & \multicolumn{6}{c}{CPU} & \multicolumn{12}{c}{GPU} \\
  \cmidrule(lr){2-7} \cmidrule(lr){8-19}
  {\#V} &      LM &      LF &      LS &    Trav &    Rep &   Total &   CtD &  LLK &  LFK &  LSK &  LEK &  CaK & SFK &   BtH &  OSK & Scan &   TwC & Total \\
      
  \midrule
  1000000  &   917.0 &   256.0 &   333.8 &   361.1 &   43.8 &  1911.8 &   7.0 &  0.4 &  0.2 &  0.2 &  1.2 &  0.4 & 0.2 &   8.8 &  0.3 &  0.3 &  19.0 &   3.2 \\
  2353515  &  2202.2 &   608.6 &   786.3 &   851.6 &  105.6 &  4554.3 &  16.2 &  1.1 &  0.6 &  0.6 &  4.5 &  0.9 & 0.5 &  20.6 &  0.6 &  0.4 &  45.8 &   9.1 \\
  3707030  &  3481.0 &   950.6 &  1240.0 &  1337.6 &  167.7 &  7177.0 &  25.4 &  1.7 &  0.9 &  0.9 &  7.5 &  1.4 & 0.7 &  31.8 &  1.0 &  0.7 &  72.0 &  14.8 \\
  5060545  &  4783.0 &  1304.8 &  1696.7 &  1834.0 &  232.0 &  9850.6 &  34.6 &  2.3 &  1.2 &  1.2 & 10.6 &  2.0 & 1.0 &  45.5 &  1.3 &  0.9 & 100.5 &  20.5 \\
  6414060  &  6046.3 &  1648.7 &  2136.4 &  2314.1 &  292.8 & 12438.3 &  43.9 &  2.9 &  1.5 &  1.5 & 13.7 &  2.5 & 1.2 &  56.8 &  1.7 &  1.1 & 126.8 &  26.1 \\
  7767575  &  7341.3 &  1994.7 &  2588.9 &  2823.8 &  353.5 & 15102.2 &  52.9 &  3.5 &  1.8 &  1.8 & 16.8 &  3.0 & 1.5 &  71.3 &  2.0 &  1.2 & 155.9 &  31.7 \\
  9121090  &  8691.1 &  2355.0 &  3052.4 &  3318.1 &  419.2 & 17835.8 &  62.0 &  4.2 &  2.2 &  2.1 & 19.9 &  3.6 & 1.7 &  83.6 &  2.4 &  1.4 & 183.0 &  37.4 \\
  10474605 &  9979.0 &  2687.9 &  3511.3 &  3802.3 &  475.2 & 20455.7 &  71.3 &  4.8 &  2.5 &  2.4 & 22.9 &  4.1 & 1.9 &  96.7 &  2.7 &  1.6 & 210.9 &  42.9 \\
  11828120 & 11246.7 &  3037.6 &  3943.3 &  4274.7 &  538.4 & 23040.8 &  80.7 &  5.4 &  2.8 &  2.7 & 25.8 &  4.6 & 2.2 & 100.0 &  3.1 &  1.8 & 229.1 &  48.4 \\
  13181635 & 12592.3 &  3381.6 &  4417.0 &  4766.7 &  606.7 & 25764.4 &  89.7 &  6.0 &  3.1 &  3.0 & 28.8 &  5.1 & 2.4 & 113.7 &  3.4 &  2.0 & 257.3 &  53.9 \\
  14535150 & 13883.2 &  3733.7 &  4872.0 &  5333.8 &  671.6 & 28494.4 &  98.5 &  6.6 &  3.4 &  3.4 & 31.9 &  5.7 & 2.7 & 124.7 &  3.8 &  2.1 & 282.7 &  59.5 \\
  15888665 & 15202.4 &  4064.6 &  5318.9 &  5699.3 &  724.6 & 31009.8 & 107.8 &  7.2 &  3.8 &  3.6 & 34.9 &  6.2 & 2.9 & 131.3 &  4.1 &  2.3 & 304.1 &  65.1 \\
  17242180 & 16545.0 &  4482.0 &  5779.2 &  6302.0 &  798.5 & 33906.7 & 117.0 &  7.8 &  4.1 &  4.0 & 38.0 &  6.7 & 3.1 & 151.2 &  4.5 &  2.5 & 338.9 &  70.7 \\
  18595695 & 17832.9 &  4778.3 &  6216.1 &  6722.0 &  856.6 & 36405.9 & 126.0 &  8.4 &  4.4 &  4.3 & 41.1 &  7.3 & 3.4 & 155.2 &  4.8 &  2.7 & 357.5 &  76.3 \\
  19949210 & 19130.3 &  5143.9 &  6667.7 &  7222.8 &  916.7 & 39081.3 & 135.4 &  9.1 &  4.7 &  4.6 & 44.2 &  7.8 & 3.6 & 168.0 &  5.1 &  2.9 & 385.3 &  82.0 \\
  21302725 & 20425.9 &  5494.6 &  7109.7 &  7707.9 &  978.0 & 41716.1 & 144.5 &  9.7 &  5.0 &  4.9 & 47.3 &  8.3 & 3.9 & 180.3 &  5.5 &  3.1 & 412.5 &  87.6 \\
  22656240 & 21727.2 &  5817.2 &  7555.6 &  8263.2 & 1046.8 & 44410.1 & 153.7 & 10.3 &  5.3 &  5.2 & 50.3 &  8.8 & 4.1 & 195.7 &  5.8 &  3.3 & 442.5 &  93.2 \\
  24009755 & 23022.5 &  6166.4 &  7999.8 &  8670.9 & 1101.9 & 46961.4 & 163.0 & 10.9 &  5.7 &  5.5 & 53.3 &  9.4 & 4.4 & 220.6 &  6.2 &  3.5 & 482.4 &  98.7 \\
  25363270 & 24414.3 &  6500.9 &  8449.3 &  9124.3 & 1178.0 & 49666.8 & 171.6 & 11.5 &  6.0 &  5.8 & 56.4 &  9.9 & 4.6 & 212.8 &  6.5 &  3.6 & 488.7 & 104.3 \\
  26716785 & 25813.9 &  6828.1 &  8889.6 &  9664.0 & 1237.3 & 52432.9 & 181.1 & 12.1 &  6.3 &  6.1 & 59.4 & 10.4 & 4.8 & 230.9 &  6.9 &  3.8 & 521.9 & 109.9 \\
  28070300 & 26987.9 &  7188.0 &  9346.4 & 10162.0 & 1293.3 & 54977.6 & 190.1 & 12.7 &  6.6 &  6.4 & 62.5 & 10.9 & 5.1 & 250.5 &  7.2 &  4.0 & 556.0 & 115.4 \\
  29423815 & 28639.1 &  7563.8 &  9806.0 & 10631.2 & 1377.3 & 58017.4 & 199.4 & 13.3 &  6.9 &  6.7 & 65.5 & 11.5 & 5.3 & 249.1 &  7.6 &  4.2 & 569.4 & 121.0 \\
  30777330 & 29646.2 &  7923.9 & 10244.3 & 11172.6 & 1427.1 & 60414.0 & 208.6 & 13.9 &  7.3 &  7.0 & 68.6 & 12.0 & 5.6 & 260.7 &  7.9 &  5.2 & 596.7 & 127.4 \\
  32130845 & 31144.5 &  8219.6 & 10730.3 & 11698.1 & 1484.5 & 63277.0 & 218.3 & 14.6 &  7.6 &  7.3 & 71.7 & 12.5 & 5.8 & 280.9 &  8.3 &  4.5 & 631.4 & 132.3 \\
  33484360 & 32300.9 &  8590.3 & 11173.1 & 12165.1 & 1549.6 & 65779.0 & 227.3 & 15.2 &  7.9 &  7.6 & 74.9 & 13.0 & 6.0 & 281.0 &  8.6 &  4.7 & 646.1 & 137.9 \\
  34837875 & 33777.4 &  8967.6 & 11631.6 & 12646.8 & 1607.7 & 68631.0 & 236.1 & 15.8 &  8.2 &  7.9 & 77.9 & 13.6 & 6.3 & 313.9 &  9.0 &  4.9 & 693.6 & 143.6 \\
  36191390 & 35064.4 &  9333.6 & 12107.0 & 13150.9 & 1672.9 & 71328.8 & 244.7 & 16.4 &  8.5 &  8.2 & 81.1 & 14.1 & 6.5 & 296.9 &  9.3 &  5.1 & 690.9 & 149.2 \\
  37544905 & 36311.0 &  9657.9 & 12577.5 & 13583.7 & 1729.4 & 73859.4 & 254.2 & 17.0 &  8.9 &  8.5 & 84.2 & 14.5 & 6.8 & 311.1 &  9.7 &  5.3 & 720.0 & 154.7 \\
  38898420 & 37995.3 &  9971.2 & 13059.2 & 14116.6 & 1796.1 & 76938.4 & 263.8 & 17.6 &  9.2 &  8.8 & 87.2 & 15.2 & 7.0 & 340.6 & 10.0 &  5.4 & 764.8 & 160.4 \\
  40251935 & 39183.2 & 10352.9 & 13479.0 & 14752.4 & 1870.0 & 79637.6 & 272.6 & 18.2 &  9.5 &  9.1 & 90.2 & 15.7 & 7.3 & 363.6 & 10.4 &  5.6 & 802.3 & 166.0 \\
  41605450 & 40622.4 & 10776.6 & 13917.1 & 15153.5 & 1920.3 & 82389.9 & 281.8 & 18.9 &  9.8 &  9.4 & 93.3 & 16.2 & 7.5 & 366.7 & 10.7 &  5.8 & 820.1 & 171.6 \\
  42958965 & 41531.3 & 11131.8 & 14290.0 & 15633.5 & 1981.9 & 84568.5 & 291.3 & 19.4 & 10.1 &  9.7 & 96.2 & 16.7 & 7.7 & 406.0 & 11.1 &  6.0 & 874.3 & 177.0 \\
  44312480 & 42827.9 & 11451.2 & 14751.8 & 16122.6 & 2046.5 & 87200.0 & 299.9 & 20.0 & 10.4 & 10.0 & 99.2 & 17.2 & 8.0 & 419.4 & 11.4 &  6.2 & 901.7 & 182.4 \\
  \bottomrule
  
  \end{tabular}
  }
  \caption{Table random meshes. The times are in miliseconds. The table presents the number of vertices (\#V) for each mesh, along with the timings for different stages of the algorithm. For the CPU version: "Label the longest-edge" (LM), "Label frontier edges" (LF), "Label seed edges" (LS), "Traversal phase" (Trav), and "Repair phase" (Rep), cumulating in the total time for CPU Polylla (Total). The GPU version encompasses: "Copy to Device" (CtD), "Label the longest-edge kernel" (LLK), "Label frontier edges kernel" (LFK), "Label seed edges kernel" (LSK), "Label extra seed and frontier edges kernel" (LEK), "Change attributes" (CaK), "Search frontier edges for each seed edge" (SFK), "Overwrite seeds" (OSK), "Scan and compact seed edges array" (Scan), "Copy back to Host" (BtH), culminating in the total time for GPU Polylla excluding copy times (Total).}\label{table:table_random_Total}
\end{table*}

\begin{table*}
  
  \resizebox{1\columnwidth}{!}{%
\begin{tabular}{lcccccccccccccccccc}
  \toprule
  {} & \multicolumn{6}{c}{CPU} & \multicolumn{12}{c}{GPU} \\
  \cmidrule(lr){2-7} \cmidrule(lr){8-19}
  \#V &      LM &      LF &      LS &    Trav & Rep &    Total &   CtD &  LLK &  LFK &  LSK &  LEK &  CaK &  SFK &    BtH &  OSK & Scan &    TwC & Total \\
  \midrule
  1000000   &   675.5 &   178.8 &   274.0 &   303.8 & 0.0 &   1432.0 &   7.1 &  0.4 &  0.2 &  0.2 &  0.4 &  0.3 &  0.2 &    9.0 &  0.2 &  0.3 &   18.3 &   2.2 \\
  4194304   &  2875.3 &   757.7 &  1164.6 &  1287.4 & 0.0 &   6084.9 &  28.6 &  1.4 &  1.0 &  1.1 &  1.3 &  1.6 &  0.7 &   37.7 &  0.8 &  0.8 &   75.1 &   8.8 \\
  7387524   &  5044.4 &  1345.6 &  2036.3 &  2293.3 & 0.0 &  10719.5 &  50.4 &  2.5 &  1.8 &  1.9 &  2.3 &  2.8 &  1.2 &   65.4 &  1.3 &  1.3 &  131.0 &  15.2 \\
  10582009  &  7315.4 &  1926.1 &  2905.5 &  3265.4 & 0.0 &  15412.4 &  73.3 &  3.6 &  2.6 &  2.6 &  3.3 &  4.0 &  1.7 &   94.8 &  1.9 &  1.8 &  189.6 &  21.6 \\
  13771521  &  9571.4 &  2501.0 &  3793.2 &  4265.4 & 0.0 &  20131.0 &  93.8 &  4.6 &  3.3 &  3.4 &  4.3 &  5.2 &  2.2 &  123.7 &  2.5 &  2.3 &  245.3 &  27.9 \\
  16966161  & 11605.6 &  3096.9 &  4678.0 &  5261.5 & 0.0 &  24642.0 & 115.3 &  5.7 &  4.1 &  4.2 &  5.3 &  6.4 &  2.7 &  149.7 &  3.1 &  2.7 &  299.2 &  34.2 \\
  20160100  & 13855.5 &  3737.9 &  5710.1 &  6318.6 & 0.0 &  29622.0 & 136.7 &  6.7 &  4.9 &  4.9 &  6.3 &  7.6 &  3.3 &  183.9 &  3.7 &  3.2 &  361.2 &  40.5 \\
  23357889  & 16018.7 &  4263.5 &  6465.5 &  7261.1 & 0.0 &  34008.7 & 158.3 &  7.7 &  5.6 &  5.7 &  7.3 &  8.9 &  3.8 &  213.3 &  4.2 &  3.7 &  418.6 &  47.0 \\
  26553409  & 18215.8 &  4855.4 &  7341.4 &  8241.9 & 0.0 &  38654.5 & 179.7 &  8.8 &  6.4 &  6.5 &  8.3 & 10.1 &  4.3 &  237.9 &  4.8 &  4.2 &  470.9 &  53.3 \\
  29746116  & 20383.7 &  5426.4 &  8232.4 &  9217.4 & 0.0 &  43259.9 & 201.3 &  9.8 &  7.2 &  7.2 &  9.2 & 11.3 &  4.8 &  269.6 &  5.4 &  4.7 &  530.6 &  59.6 \\
  32936121  & 22671.5 &  6017.1 &  9151.4 & 10215.1 & 0.0 &  48055.0 & 223.3 & 10.8 &  7.9 &  8.0 & 10.3 & 12.5 &  5.3 &  299.9 &  6.0 &  5.2 &  589.2 &  65.9 \\
  36132121  & 24764.4 &  6584.7 &  9991.2 & 11257.4 & 0.0 &  52597.7 & 244.6 & 11.9 &  8.6 &  8.8 & 11.2 & 13.7 &  5.8 &  327.6 &  6.5 &  5.6 &  644.3 &  72.2 \\
  39325441  & 27005.2 &  7180.2 & 10944.8 & 12258.3 & 0.0 &  57388.5 & 266.1 & 12.9 &  9.5 &  9.6 & 12.2 & 15.0 &  6.3 &  357.2 &  7.1 &  6.1 &  701.9 &  78.7 \\
  42510400  & 29660.8 &  7785.0 & 11785.1 & 13269.7 & 0.0 &  62500.6 & 288.6 & 13.9 & 10.2 & 10.4 & 13.2 & 16.2 &  6.8 &  380.2 &  7.7 &  7.6 &  754.7 &  85.9 \\
  45711121  & 31443.8 &  8379.1 & 12655.1 & 14243.5 & 0.0 &  66721.5 & 310.6 & 14.9 & 11.0 & 11.1 & 14.2 & 17.3 &  7.3 &  405.0 &  8.3 &  7.1 &  806.8 &  91.2 \\
  48902049  & 33451.3 &  8912.5 & 13485.0 & 15180.5 & 0.0 &  71029.3 & 332.1 & 15.9 & 11.8 & 11.9 & 15.1 & 18.6 &  7.8 &  436.8 &  8.9 &  7.5 &  866.5 &  97.6 \\
  52099524  & 35681.5 &  9521.8 & 14447.9 & 16203.7 & 0.0 &  75854.9 & 351.8 & 17.0 & 12.6 & 12.7 & 16.1 & 19.8 &  8.3 &  455.3 &  9.4 &  8.0 &  911.1 & 104.0 \\
  55294096  & 37863.3 & 10056.3 & 15205.1 & 17319.7 & 0.0 &  80444.3 & 375.2 & 18.0 & 13.3 & 13.5 & 17.1 & 21.0 &  8.8 &  519.1 & 10.0 &  8.5 & 1004.6 & 110.3 \\
  58476609  & 39853.3 & 10606.7 & 16053.8 & 18156.3 & 0.0 &  84670.1 & 397.0 & 19.0 & 14.2 & 14.4 & 18.0 & 22.3 &  9.3 &  554.4 & 10.6 &  9.0 & 1068.1 & 116.7 \\
  61669609  & 42131.3 & 11209.5 & 17008.9 & 19147.3 & 0.0 &  89497.0 & 445.2 & 20.0 & 15.0 & 15.1 & 19.1 & 23.5 &  9.8 &  580.1 & 11.2 &  9.5 & 1148.3 & 123.1 \\
  64866916  & 44582.9 & 11927.8 & 17936.0 & 19990.2 & 0.0 &  94436.9 & 439.3 & 21.0 & 15.8 & 15.9 & 20.0 & 24.7 & 10.3 &  708.9 & 11.7 &  9.9 & 1277.6 & 129.4 \\
  68062500  & 46971.4 & 12419.7 & 18752.5 & 21187.0 & 0.0 &  99330.5 & 462.9 & 22.1 & 16.5 & 16.7 & 21.0 & 25.9 & 10.8 &  806.8 & 12.3 & 10.4 & 1405.5 & 135.8 \\
  71250481  & 49320.1 & 13090.8 & 19816.6 & 22347.2 & 0.0 & 104574.6 & 495.3 & 23.0 & 17.3 & 17.4 & 21.9 & 27.2 & 11.3 &  847.5 & 12.9 & 10.9 & 1484.8 & 142.0 \\
  74459641  & 51456.9 & 13746.1 & 20822.9 & 23551.0 & 0.0 & 109577.0 & 522.0 & 24.1 & 18.1 & 18.2 & 22.9 & 28.4 & 11.8 &  876.9 & 13.5 & 11.4 & 1547.2 & 148.3 \\
  77651344  & 53773.7 & 14553.1 & 22019.6 & 24652.1 & 0.0 & 114998.4 & 536.4 & 25.1 & 18.9 & 19.1 & 23.9 & 29.6 & 12.3 &  912.7 & 14.1 & 11.8 & 1603.9 & 154.8 \\
  80838081  & 55984.1 & 15378.2 & 23099.2 & 25122.1 & 0.0 & 119583.8 & 550.0 & 26.1 & 19.7 & 19.8 & 24.9 & 30.8 & 12.8 &  848.5 & 14.6 & 12.3 & 1559.5 & 161.1 \\
  84033889  & 57761.7 & 15477.8 & 23321.8 & 26418.4 & 0.0 & 122979.6 & 584.3 & 27.1 & 20.5 & 20.6 & 25.8 & 32.0 & 13.3 &  872.2 & 15.2 & 12.8 & 1623.7 & 167.3 \\
  87216921  & 60788.6 & 16596.3 & 24310.7 & 27955.6 & 0.0 & 129651.2 & 598.4 & 28.1 & 21.2 & 21.4 & 26.8 & 33.2 & 13.8 &  956.1 & 15.8 & 13.3 & 1728.0 & 173.6 \\
  90421081  & 62680.7 & 16551.4 & 25045.8 & 28437.6 & 0.0 & 132715.4 & 742.1 & 29.1 & 22.1 & 22.2 & 27.8 & 34.5 & 14.3 &  938.6 & 16.4 & 13.8 & 1860.7 & 180.0 \\
  93605625  & 64657.3 & 17209.7 & 25838.4 & 29600.6 & 0.0 & 137305.8 & 719.3 & 30.1 & 22.8 & 23.0 & 28.7 & 35.7 & 14.8 &  986.0 & 16.9 & 14.2 & 1891.6 & 186.3 \\
  96805921  & 66426.1 & 17874.2 & 27100.4 & 30790.0 & 0.0 & 142190.6 & 665.5 & 31.1 & 23.6 & 23.7 & 29.6 & 36.9 & 15.3 & 1020.5 & 17.5 & 14.7 & 1878.5 & 192.5 \\
  100000000 & 68806.4 & 18513.8 & 27816.8 & 31814.4 & 0.0 & 146951.4 & 741.0 & 32.1 & 24.4 & 24.6 & 30.3 & 38.1 & 15.8 & 1027.5 & 18.1 & 15.2 & 1967.0 & 198.4 \\
  \bottomrule
  \end{tabular}
  }

  \caption{Table grid. The times are in miliseconds. The table presents the number of vertices (\#V) for each mesh, along with the timings for different stages of the algorithm. For the CPU version: "Label the longest-edge" (LM), "Label frontier edges" (LF), "Label seed edges" (LS), "Traversal phase" (Trav), and "Repair phase" (Rep), cumulating in the total time for CPU Polylla (Total). The GPU version encompasses: "Copy to Device" (CtD), "Label the longest-edge kernel" (LLK), "Label frontier edges kernel" (LFK), "Label seed edges kernel" (LSK), "Label extra seed and frontier edges kernel" (LEK), "Change attributes" (CaK), "Search frontier edges for each seed edge" (SFK), "Overwrite seeds" (OSK), "Scan and compact seed edges array" (Scan), "Copy back to Host" (BtH), culminating in the total time for GPU Polylla excluding copy times (Total).}\label{table:table_grid_Total}
\end{table*}

\end{document}